\documentclass{jov}     
\usepackage{natbib}
\usepackage{comment}
\usepackage[T1]{fontenc}
\usepackage{makecell}
\usepackage{graphicx}
\usepackage{subcaption}
\usepackage{wrapfig}
\usepackage{fixltx2e}
\usepackage{amssymb}
\usepackage{gensymb}

\begin{document}

\title{Recent consumer OLED monitors can be suitable for vision science}
\abstract{Vision science imposes rigorous requirements for the design and execution of psychophysical studies and experiments. These requirements ensure precise control over variables, accurate measurement of perceptual responses, and reproducibility of results, which are essential for investigating visual perception and its underlying mechanisms. Since different experiments have different requirements, not all aspects of a display system are critical for a given setting. Therefore, some display systems may be suitable for certain types of experiments but unsuitable for others. An additional challenge is that the performance of consumer systems is often highly dependent on specific monitor settings and firmware behavior. Here, we evaluate the performance of four display systems: a consumer LCD gaming monitor, a consumer OLED gaming monitor, a consumer OLED TV, and a VPixx PROPixx projector system. To allow the reader to assess the suitability of these systems for different experiments, we present a range of different metrics: luminance behavior, luminance uniformity across display surface, estimated gamma values and linearity, channel additivity, channel dependency, color gamut, pixel response time, and pixel waveform. In addition, we exhaustively report the monitor firmware settings used. Our analyses show that current consumer-level OLED display systems are promising, and adequate to fulfill the requirements of some critical vision science experiments, allowing laboratories to run their experiments even without investing in high-quality professional display systems. For example, the tested Asus OLED gaming monitor shows excellent response time, a sharp square waveform even at $240\; Hz$, a color gamut that covers $94\%$ of DCI-P3 color space, and the best luminance uniformity among all four tested systems, making it a favorable option on price-to-performance ratio.}

\author{Abu Haila}{Tarek}
 {Centre for Cognitive Science, Institute of Psychology}
 {Technical University of Darmstadt, Darmstadt, Germany}
 {https://www.psychologie.tu-darmstadt.de/perception/home_per/index.en.jsp}{Tarek.Haila@tu-darmstadt.de}
\author{Kunst}{Korbinian}
 {Laboratory of Adaptive Lighting Systems and Visual Processing}
 {Technical University of Darmstadt, Darmstadt, Germany}
 {https://www.lichttechnik.tu-darmstadt.de/fachgebiet_lichttechnik_lt/index.en.jsp}{kunst@lichttechnik.tu-darmstadt.de}
 \author{Khanh}{Tran Quoc}
 {Laboratory of Adaptive Lighting Systems and Visual Processing}
 {Technical University of Darmstadt, Darmstadt, Germany}
 {https://www.lichttechnik.tu-darmstadt.de/fachgebiet_lichttechnik_lt/index.en.jsp}{office@lichttechnik.tu-darmstadt.de}
  \author{Wallis}{Thomas S. A. }
 {Centre for Cognitive Science, Institute of Psychology}
 {Technical University of Darmstadt, Darmstadt, Germany}
 {https://www.psychologie.tu-darmstadt.de/perception/home_per/index.en.jsp}{Thomas.Wallis@tu-darmstadt.de}

\keywords{vision science, color, luminance, perception, display, calibration}

\maketitle

\begin{figure}
    \centering
    \includegraphics[width=0.7\textwidth]{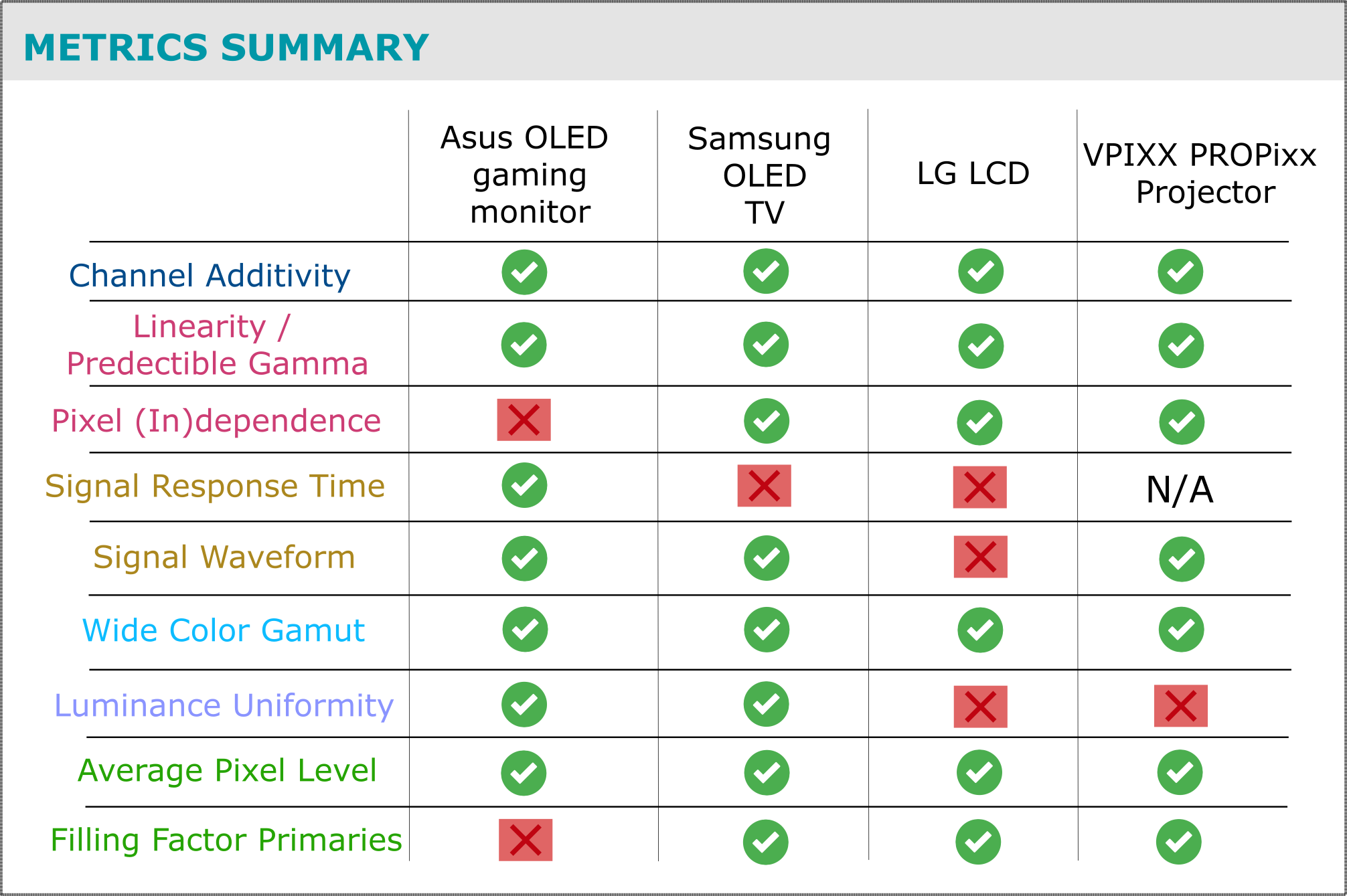}
    \caption{
    A summary of the performance of the four tested display systems on some of the display properties we consider.
    The pass-fail indicators should not be interpreted as definitive criteria for the acceptance or rejection of any particular system. Rather, they serve as a means to evaluate whether a system meets its theoretical hardware specifications and operates within expected performance limits.}
    \label{fig:teaser}
\end{figure}

\section{Introduction}
A pillar of vision science is a display system that is both accurate and stable (``Know Thy Stimulus''; \citep{geisler1987ideal, Wichmann_Jaekel_2018}). 
However, since different experiments have different requirements, not all aspects of a display system are critical for a given setting. Therefore, different characteristics could be of more or less importance than others depending on the requirements of the experiment at hand. 

The Cathode-Ray Tube (CRT) display has long been considered the ``gold standard'' for many experiments. Consumer liquid-crystal displays (LCD), for example, are still considered unsuitable for many high-precision experiments. 
Briefly, CRT displays operate using an electron beam in a vacuum and a phosphor screen, by modulating the electron beam the intensity, hence the luminance of the output is controlled. Color CRTs use three beams, one for each of the RGB primaries, and different phosphor coatings that help in producing the RGB colors. To display a frame, the beams need to project and scan the surface of the screen line by line from top to bottom, left to right \citep{Wheeler199_CRT_tech}.
In LCD displays, the liquid crystal molecules are placed behind each pixel and change orientation, in synchrony or individually, to control the output light. The light in LCD comes from what is known as a \textit{backlight}, usually white, and is filtered when passing through RGB filters that constitute each pixel to output the desired color \citep{Fujieda.ch5, Fujieda.ch2}.

\citet{Elze2012TemporalPO} discussed various aspects and shortcomings of the LCD technology of that time. They addressed the problem of motion blur in LCDs and the luminance stepping in which the different primaries (R, G, and B) show different response time profiles when making the transition from 0 to 255 (8-bit) due to the different driving signals to the different primaries. They tested, as well, the response time variability that is reported to be in the range up to $10 \;ms$ and the variation coefficient (relative deviation with respect to the absolute response time calculated by dividing the standard deviation by the mean) amounting to 0.25 on average. These reported values are considered unsatisfactory for many vision science experiments.

CRT displays have, as well, their share of problems. \citet{Lagroix2012LCDsAB} pointed out how CRT displays actually suffer from \textit{display persistence}, residual image, while the tested LCD counterparts showed no such effect. Residual image in CRT displays is the result of the phosphor persistence that remains visible for some time after the image has been turned off. They made it clear that concerning the response time of a pixel (rise and fall) LCD displays have come a long way thanks to overdrive technology\footnote{Overdrive helps in speeding up the liquid-crystal transition from one state to another (i.e. one brightness level to another) by applying higher voltage. It helps to achieve a faster response time and to eliminate the chance of the ghosting effect.}. They showed a substantial improvement by hitting $1-6\; ms$ for the rise time and less than $2\; ms$ for the fall time, beating their stigma for having a sluggish response that was reported in various papers in the year 2007 and before.
\citet{Klein2013PhotometricAC} pointed out one of the major shortcomings in discussing the inhomogeneity (i.e. non-uniformity) in luminance across CRT displays. Their measurements showed up to $10\%$ variation across the display surface, while other authors reported up to $20\%$ variation \citep{bohnsack1997characteristics, Krantz2000_tell_me}. Over the subsequent years, CRT displays have become increasingly rare, phased out, and difficult to find due to advancements in various consumer display technologies meaning that they are rarely, if ever, commercially available.

In contrast, Organic Light-Emitting Diode (OLED) display technology has become more affordable and abundant on the market in recent years. 
OLED displays are composed of self-emissive diodes that allow the control of pixels independently from one another and eliminate leakage caused normally by the use of a backlight (e.g. in LCDs). 
This means that in theory, OLED pixels can produce truly zero luminance, and also that pixels should be completely independent (switching a certain pixel on and/or off would not affect the luminance value of other pixels).
These properties of the display technology mean that the contrast range ($\frac{max\:white\: luminance}{min\: black\: luminance}$) of these monitors can approach infinity. They can also offer a larger color gamut thanks to their saturated primaries and narrow-band spectra in comparison to LCD and CRT displays. Thanks to the Helmholtz-Kohlrausch effect, more saturated colors are perceived to be brighter  \citep[i.e. an increase in color saturation increases the perceived brightness; ][]{Tsujimura_Outlook_OLED, Tsujimura_2017_ch4}. 
Since the perception of brightness is also linked to and affected by the surround colors / ambient light \citep{Bartleson:67} the same color on a darker background can be perceived as brighter. 
According to \citet{hack201060} various psychological and perception studies have reported that thanks to the OLED pitch-blackness ability, users have reported an increase in perceived brightness at the same level of luminance as their counterparts LCD displays.
Finally, previous measurements from \citet{Elze_OLED_2013} showed that OLEDs are capable of producing extremely precise, almost-square wave, temporal response profiles.
Together, these results suggest that OLED technologies hold much promise for use in vision science.

Several previous papers have investigated the suitability of OLED displays for vision science applications.
\citet{Cooper2013OLED} presented a thorough investigation of the suitability of OLED monitors for vision science research and concluded that they are a favorable option. They included in their study two OLED monitors, namely the Trimaster EL BVM-F250 Master Monitor (BVM), a high-end professional video monitor for production applications, and the Trimaster EL PVM- 2541 Picture Monitor (PVM), a video monitor designed for general usage. They showed that OLED displays can have a nice additivity property, a wide color gamut, and their gamma functions can be fitted to a power function making their luminance output linearizable if desired. One of the tested OLED showed, as well, pixel independence in the sense that the behavior of vertical pixels is not affected or influenced by horizontal pixels and vice-versa.
\citet{Elze_OLED_2013} tested a very specific medical OLED monitor, namely Sony PVM 2551MD, and whether it complies with the DICOM\footnote{Digital Imaging and Communication in Medicine.} standards. They investigated the discretization of the color channels and showed that $\approx 50\%$ of the neighboring bit values are perceptually indistinguishable (based on the just-noticeable difference (JND) defined in the DICOM standards) for the blue channel and $\approx 28\%$ for the green channel rendering the effective perceptual luminance resolution of the monitor to be considerably below it is actual 8-bit resolution. They showed as well that activating all the monitor's pixels, a full-screen color patch, results in early saturation in contrast relative to a small color patch. For instance, the white color patch showed saturation in full-screen mode as early as $\approx 162\; cd/m^2$ (corresponding to a bit value of 196) as opposed to the monotonic increase in luminance while increasing the bit value for a small circular color patch that reaches eventually $\approx 400 cd/m^2$.
In a recent study, \citet{Ashraf24} present measurements of an OLED TV using different regression models on how to run color calibration on a 4-primary OLED monitor. The monitor they tested showed a saturation effect across all its channels near the upper limit of the bit depth (bit value $> 700$ in 10-bit depth), breaking any possible linear behavior or compliance to a gamma function. 
Their work focused only on the color calibration aspect and concluded that the performance of the different color calibration models was best for low luminance levels and worst for higher luminance levels. 
\citet{Tian_OLED_HDR} showed that the Sony A1 OLED TV suffered from a drastic drop in peak luminance for all its color channels and any window size $>10\%$ under the default settings they chose, and hence they fixed the window size for all their measurements to $4\%$. 
\citet{Yang_OLED_vs_LED} showed that there is a drop in luminance for the OLED display of a DELL XPS 15 7590 notebook, when the average pixel level factor $>1\%$, a drop from $\approx 562$ to $431 \; cd/m^2$. Evaluating the PVM-2541 OLED, \citet{Ito2013OLED} showed that when the filling factor is $>40\%$ and the bit value is $>220$ (8-bit) then the luminance output drops drastically at least $\approx5\%$ and at most $\approx 27\%$. They also measured how changing the background color affects the luminance output by showing that any value above gray 220 (8-bit) for the background causes a drop in the target luminance up to $\approx 27\%$ when measuring a small square in the middle of the display, more noticeably for any gray value above $128$, but nonetheless even a gray value of 40 showed the same tendency.
Finally, \citet{dimigenHighspeedOLEDMonitor2024} recently published results evaluating a number of different monitors with a focus on the ASUS OLED panel that we also evaluate here.
We compare their results to ours more thoroughly in our Discussion section.
In general, the results of the study by \citet{dimigenHighspeedOLEDMonitor2024} agree with our measurements on the overall performance of this monitor, providing some additional confidence in the results we present here.


Several commercial display systems exist that are purpose-built for vision science, notably the VIEWPixx LCD range and/or PROPixx projector (VPixx Technologies Inc.,   Saint-Bruno, QC Canada), and the \textit{Display++} LCD range (Cambridge Research Systems Ltd., Rochester, UK). \citet{Ghodrati2015TheO} compared five high-end monitors, including those that are targeted to vision science specifically, and discussed their (un)suitability for vision science research. The tested monitors were EIZO FG2421 LCD, VIEWPixx 3D Lite LCD, Samsung 2232RZ LCD, CRS Display++ LCD, and Sony CPD-G520 CRT. They reported their findings regarding luminance perception and viewing angle, luminance uniformity, and temporal dynamics. At low luminance levels, for instance, high variability was an issue for the tested CRT. While luminance non-uniformity across the display's surfaces was as high as $20\%$ for EIZO, Samsung, and Display++ systems. All tested LCDs suffered from luminance hysteresis, in which the desired target luminance level is not achieved in a single frame when making the transition from black to white or vice-versa. The hysteresis is attributed to the backlight mechanism implemented in all tested displays. 

In this paper, we investigate the suitability of four display systems for use in vision science.
The requirements for a display device depend on the experiment; deciding which aspects are critical is part of scientific training, and not the subject of this paper.
Here, we evaluate displays on the following metrics that, we hope, paint a useful picture of how far the OLED display technology has come and how suitable it is to be deployed in vision science. 
We consider the following display system properties:
    \textbf{Channel Additivity:} defining the relationship between the response of the individual R, G and B channels and their collective response when they act in synchrony to reproduce grayscale (R=G=B).
    \textbf{Linearity / Gamma Predictability:} defining the relationship between the pixel bit value and its luminance output. Output is predictable based on the bit value, meaning that it can be corrected predictably using a lookup table or gamma function.
    \textbf{Pixel (In)Dependence:} Whether the behavior of a subset of pixels is influenced by the state of other pixels.    
    \textbf{Signal Response Time:} how long it takes a set of pixels to change state.
    \textbf{Signal Waveform:} defining the signal waveform indicating the actual duration pixels take in one state at the desired luminance level before switching to another state.
    \textbf{Color gamut:} how wide the color gamut a certain display offers (i.e. number of available colors).    
    \textbf{Luminance Uniformity:} how homogeneous the luminance output is across the display surface.
    \textbf{Average Pixel Level:} the relationship between the overall active pixels intensity (bit value) and the luminance output to determine how much the luminance output is invariant to the overall pixel values and proportion of the display using grayscale (R=G=B).
    \textbf{Filling factor primaries:} same as the average pixel level. However, instead of testing only the behavior of grayscale, we test the RGB primaries's behavior.
The display systems we test consist of two OLEDs, an LCD, and an LED projector; these are described in more detail below. Overall, we find that consumer-grade OLED technology is promising as an affordable alternative to LCDs and CRTs, that can still be as reliable and competent in important aspects of performance as some specialized display systems. Fig.\ref{fig:teaser} offers a coarse summary of each tested system's performance on the different properties we consider.

\section{Material and Methods}

\subsection{Hardware}
\label{subsec:hardware}

\subsubsection{Display Systems}
We test four displaying systems: the PROPixx projector (VPixx Technologies Inc.), ASUS OLED ROG Swift PG27AQDM gaming monitor, a Samsung OLED TV, and an LG UltraGear 27GN950-B LED gaming monitor, check Table \ref{tab:diplays_specs}.

The PROPixx projector is designed specifically for vision science applications. It is based on RGB narrow LEDs and has in theory 12-bit depth per channel. Its native resolution is $1920\times1080$ at $120 Hz$, and it can operate at frequencies of up to $480 \;Hz$ in color and up to $1440\; Hz$ in grayscale at a lower spatial resolution. The projector uses DMD technology for displaying and light manipulation, it uses the AeroView 100 screen (Stewart Filmscreen, Torrance, Canada). Its output is already perfectly linear, so gamma correction is unnecessary. The throw distance between the rim of the lens and the projection screen measured to $\approx 92.5\;cm$ and the projected image measured to $\approx 97.5\times54.5\;cm$ and displayed on the AeroView 100 rear projection screen. The measurements of the PROPixx serve as a baseline in this paper.

The ASUS ROG Swift OLED PG27AQDM monitor is 27" and uses LG OLED RGBW panel. It offers 10-bit depth per channel and coverage of up to $94\%$ of the DCI-P3 color gamut. It has a native resolution of $2560\times1440$ and can operate at a frequency of up to $240\;Hz$. The display can reach a peak brightness of $500 \: cd/m^2$ under certain settings.

The LG UltraGear 27GN950-B is 27" and uses white LED as a backlight. It has a native resolution of $3840\times2160$ (4K) with a bit depth of 10 bits and offers up to $95\%$ coverage of DCI-P3 color space. Its refresh rate can go up to $144\; Hz$. The maximum luminance of the monitor is advertised to reach up to $400 \: cd/m^2$.

Finally, the Samsung OLED TV display GQ65S93CAT is 65" and offers 10-bit depth with a native resolution of $3840\times2160$ (4K) and can be driven up to $144\; Hz$. It uses quantum-dot technology as the underlying lighting mechanism.

The difference between the two OLED displays (Asus vs. Samsung TV) is that Asus OLED uses four subpixels to reproduce colors, namely the common RGB + an additional white subpixel based on the LG panel \textit{LW270AHQ-ERG2}, which enables the display to reach higher brightness levels. The Samsung OLED TV, on the other hand, uses a blue OLED backlight and quantum-dot technology to produce pure and saturated RGB primaries, which allows the display to have a noticeably larger color gamut and more saturated colors. Quantum-Dots (QD) have the properties of narrow and brighter emission, a higher signal-to-noise ratio thanks to their inorganic composition in comparison to organic dyes. One major difference is that the full width at half maximum (FWHM) of the emission peak is as half as that of organic LED, $20-30\; nm$ vs. $50\;nm$. Hence, better color contrast and saturation \citep{QD_Applications}.

The pixel density of each display in millimeters is calculated by dividing the diagonal pixels over the diagonal physical length in millimeters as shown by Eq.\ref{eq:ppm} i.e. for ASUS OLED ROG Swift at a resolution of $2560\times1440 \rightarrow \sqrt{2560^2 + 1440^2} = 2937.2$ pixels diagonally, and the effective surface area measures to approximately $587\times329.5\; mm$ and hence diagonally it is $673.1\; mm$ or as advertised $26.5 \; in$. This gives $\frac{2937.2}{673.1} = 4.36 \; pixels/mm$. The pixel density besides some other main characteristics of each of the displays are summarized in Table \ref{tab:diplays_specs}.

\begin{equation}
    PPM = \frac{dp}{dl} = \frac{\sqrt{X^2 + Y^2}}{W^2 + H^2}
\label{eq:ppm}
\end{equation}
where $PPM$ is pixels per millimeter, $dp$ is diagonal pixels, $dl$ is diagonal length in millimeters, $X$ is horizontal pixels, $Y$ is vertical pixels, $W$ is width in $mm$ and $H$ is height in $mm$.

\subsubsection{Measuring Devices}
For measurements, we used the following instruments: A colorimeter from X-Rite (i1Display) for calibrating the monitors and measuring the luminance levels $(cd/m^2)$. The colorimeter luminance range is $0.1 - 1000\; cd/m^2$. An X-Rite spectrophotometer (i1Pro 3) for measuring the spectral power distribution curves with operating range of $380-730\;nm$ and optical resolution of $10\;nm$. A response box (RB) from OSRTT (OSRTT Pro CS) to measure luminance transition response time and pixels waveform. The RB comprises 6 photodiode sensors aligned horizontally and occupies a width of $ \approx 19\;mm$.

We verified the accuracy and reliability of the colorimeter and the spectrophotometer measurements against a more precise and calibrated instrument, namely a spectroradiometer from Konika Minolta (CS2000). The minimum sensitivity of the CS2000 is $0.003\;cd/m^2$ at $1^{\circ}$. OLED displays are capable of switching off their pixels completely, as mentioned earlier, to represent black, as well as their capabilities of representing very low luminance levels. As a result, very low luminance levels, $<0.1\; cd/m^2$, were not possible to register using the colorimeter (i1Display) even though some of these luminance levels could be perceptible by the human eye after good dark adaptation.

The measurements for the Asus and LG displays were performed on Ubuntu 22.04 and a control PC with Intel Core i9 14th Gen., 32 GB RAM, and AMD Radeon RX 7800 XT 16 GB. The control PC for the Samsung display was an Intel Core i9 13th Gen. 64 GB RAM, NVidia GeForce RTX 4090, and under Ubuntu 20.04.
For PROPixx, the driving PC was an Intel Core i7 11th Gen., 16 GB RAM, and AMD RX6600 XT 8GB, under Ubuntu 22.04.
Only the signal response/waveform measurements were measured under Windows OS 11 because the used measuring instrument, \textit{OSRTT Pro CS}, runs exclusively on Windows.

Before every measurement session, we checked the system calibration profile for any deviation in peak luminance, gamma, or color accuracy and we re-calibrated to the target settings if needed. 
We left the display to stabilize before recording any measurements at least for $30-45$ minutes.

Due to the resolution of the X-Rite i1Display colorimeter (minimum sensitivity that starts at $0.1\; cd/m^2$), there were some unregistered luminance values at the beginning of the bit range for the OLED displays that were too dim to be registered by the colorimeter. These values were omitted.

\begin{table}[]
\footnotesize
    \centering
    \caption{A summary of the display system's main specifications.}
    \label{tab:diplays_specs}
\begin{tabular}{|l|l|l|l|l|l|l|}
\hline
\thead{Display} & \thead{Model} & \thead{Light Technology}& \thead{Native Resolution} & \thead{pixel density per mm} & \thead{Bit Depth} & \thead{Max. Refresh rate} \\ \hline
Asus & ROG Swift PG27AQDM & OLED & $2560\times1440$ & 4.36 & 10 & 240 \\      \hline
LG & UltraGear 27GN950-B & White LED & $3840\times2160$ & 6.48 & 10 & 144 \\  \hline
Samsung & GQ65S93CAT & OLED & $3840\times2160$ & 2.70 & 10 & 144 \\           \hline
PROPixx Projector & VPX-PRO-5050C & RGB LED & $1920\times1080$ & 1.95 & 12 & 1440 \\ \hline
\end{tabular}
\end{table}

%

\subsection{Metrics}
In our test, we devised the following metrics that would offer a good insight into the capability and the suitability of the displaying systems under the test. 

\subsubsection{Luminance Response}
The luminance ramp is the transition behavior of each channel along the available bit depth. We measure each of the primaries (R, G, and B) as well as the grayscale (i.e. R=G=B). All the luminance measurements are taken from a small patch ($512\times512$ pixels) in the center of the display against a mid-gray background unless stated otherwise. The luminance transition covers the whole available bit depth (10 or 12 bits) depending on the capabilities of the tested display. The measurements were done in bit steps of one or five. Five-step transition was used only for the 12-bit depth of PROPixx because the measurements were time-consuming (even with five-step transition one-channel measurement would still take up to 50 minutes). All other measurements were done at one-step transition.

\subsubsection{Pixel (In)dependence}
Pixel (in)dependency is a measure to check how the luminance output of a certain group of pixels would be affected by the behavior of another group of pixels at a different region of the display. We verified this in two ways: 1- While keeping the measured color patch at the center and only changing the background color from mid-gray to black. 2- measuring a small patch in the center on a mid-gray background while fixing two primaries to the maximum while varying the third one e.g. $[R_{max}, G_{max}, B_{DDL}]$ (DDL: Digital Driving Level or bit value). We referred to the second scenario as \textit{Other Channels Saturated (OCS)}.

\subsubsection{Linearizablitiy and gamma predictability}
A linear response of a system is a very important property for various psychophysical experiments, for a linear pixel response ensures that a generated stimulus at a certain luminance level is capable of  generating double the luminance when the pixel value of that stimulus doubles. All displays have what is known as a \textit{gamma function ($\gamma$)} which is, in its simplest form, a non-linear power function that relates between the input value ($V_{in}$) and the output value ($V_{out}$) Eq.\ref{eq:gamma_func} \citep{gamma_vision.ch.5, rehab_gamma, Pelli91}.

\begin{equation}
    V_{out} = V_{in}^\gamma 
\label{eq:gamma_func}
\end{equation}

A good display system would either show a perfect linear behavior or a predictable and steady gamma behavior giving the possibility for linearization upon decoding the applied gamma value. Because the PROPixx projector is designed with vision science in mind, it comes with a perfect linear behavior for all its channels (RGB). However, commercial displays come with various gamma functions depending on the purpose of viewing (e.g. cinema, bright surround, dark surround, HDR, gaming...etc.).

In case of a deviation in the gamma encoding values across the individual RGB channels, it is advised to control and decode each channel separately with its precise value to ensure a perfectly linear behavior of each channel. The encoded gamma behavior of each channel is determined by the ICC calibration profile, besides the behavior of the display hardware itself.

\subsubsection{Channel Additivity}
Additivity checks whether the response of the individual RGB channels sums up to the response of the channels when they act collectively, Eq.\ref{eq:RGB_summation}, to produce a grayscale (black to white). Hence, a channel output is invariant and independent of the state of other channels, at the same time it has the same output when combined with other channels to act in synchrony. 
\begin{equation}
    R_n+G_n+B_n \stackrel{?}{=} {(RGB)}_n \;\; \forall \;\; n \in [0 - (2^{bit}-1)]
\label{eq:RGB_summation}
\end{equation}

\subsubsection{Luminance Uniformity}

Spatial uniformity is a metric that ensures that the luminance output is invariant to its location on the display surface. We divide the displays into a $3\times3$ grid and measure the grayscale of each section individually. 

\subsubsection{Color Gamut}
Color gamut is defined, usually, by a triangle on CIE 1931 xy-Chromaticity diagram, the tips of this triangle are the xy-Chromaticity coordinates of each of a display's RGB primaries at their maximum and acting separately \citep{colorGamut.ch6, colorGamut.ch9}.
Each color gamut also has a a white-point that defines the chromaticity of the color \textit{White}. It determines how all other colors are rendered relative to this CCT \citep{principles_of_colors.ch7}. A common white-point for displays is a D65 (6500 \textit{Kelvin}).

Keep in mind that a color gamut is a 3D representation of the colors, whereas CIE 1931 diagram is only 2D with the assumption of fixing the luminance level to a certain point. Hence, the luminance dimension is missing. Both the luminance capabilities and the spread of the primaries define how large and rich a color gamut is, along with the usable bit-depth defined by the display.

\subsubsection{Spectral Power Distribution}
The spectral power distribution (SPD) defines the spectral profile of the light sources used in a display along the visible domain, usually $380 - 780\; nm$, and paints a picture of how efficient these primaries are at mixing colors which defines the size of the color gamut eventually.
Narrower spectral curves mean more saturated primaries, corresponding to a larger color gamut.
For these measurements, we used the X-Rite i1Pro 3 spectrophotometer.

\subsubsection{Signal Response Time and Waveform}
The response time is measured for each display using the Gray-to-Gray concept, which measures the time it takes pixels to transition from one grayscale value to another \citep{ICDM.ch10}.

We measured the waveform of exactly 1-frame and 11-frames, transitioning from black to white $(0.0 \rightarrow 1.0)$, within a second for each of the displays.
The duration of a frame is a function of the refresh rate, $\frac{Time}{Refresh\; Rate}$. For example, for a 240 Hz display, a frame should last $1000 / 240 = 4.16 \; ms$. A good waveform would be an instantaneous rise and fall of the signal on demand so that the target luminance level is reached exactly when called so as to appear exactly at the beginning of the frame and vanish with its sharp transition back at the end of the frame so one has the stimulus visible for exactly the whole frame period with no retention effect in the best case scenario.

\subsubsection{Filling Factor}
Historically people used the term \textit{Average Picture Level}, however, it has been suggested to be replaced with \textit{Average Pixel Level} (APL) for it is more accurate \citep{Poynton2012DigitalVA}. APL is expressed as a percentage of the maximum brightness of a display as a function of the averaged bit value which is the average intensity of all active pixels across a display. APL always uses a white patch on a black background for evaluation, so when $APL=0\%$ means all pixels are black, $APL=50\%$ means half the display is white, and $APL=100\%$ all pixels are white.
We analyze the grayscale in addition to each of the primary channels individually (R, G, and B). For this reason, we will use a more generic term like the \textit{Filling Factor} to express the percentage of the active pixels of a certain bit value across (R, G, B, and Grayscale) beyond being restricted only to a white patch. We will mention as well the state of the background pixels to be either black, white, or mid-gray. This thorough analysis of the different combinations will provide a detailed insight into the behavior of the primary channels concerning the average intensity of the active pixels.

Many display technologies, inherently, have a built-in safety mechanism known as \textit{Automatic Brightness Limiting} (ABL) which is triggered proportionally to the APL value to limit and protect the hardware from high power consumption, pixels aging, and heating. Some displays may offer an option that disables or limits the action of ABL, however, at the expense of the achievable peak luminance.

\subsection{Calibration}

All tested displays, except PROPixx, were color-calibrated using DisplayCal calibration software and the i1Display colorimeter before recording any of the measurements. The software allows the user to specify the parameters of calibration manually if desired. We chose to calibrate with the following parameters for the three commercial displays: max luminance: $250\; cd/m^2$ for Asus OLED \& LG LCD, $192\; cd/m^2$ for Samsung OLED TV; Color Temperature: 6500 K; Tone Curve: Gamma 2.2; Calibration Speed: High (around 16 minutes).
PROPixx offers several sequencer modes in which it controls the output color (RGB vs. grayscale) and the refresh rate (120, 180, 240, 480...etc. Hz). VPixx support advised us to choose \textit{RGB 120 Hz Calibrated High Bit Depth} sequencer to target accurate luminance representation, though this mode drops the maximum luminance to around $\approx 128\; cd/m^2$.

Full specifications of the monitor settings we used for each display are provided in the Appendix \ref{ap:display_settings}.

\section{Results}

\begin{figure}[!h]
    \centering
    \includegraphics{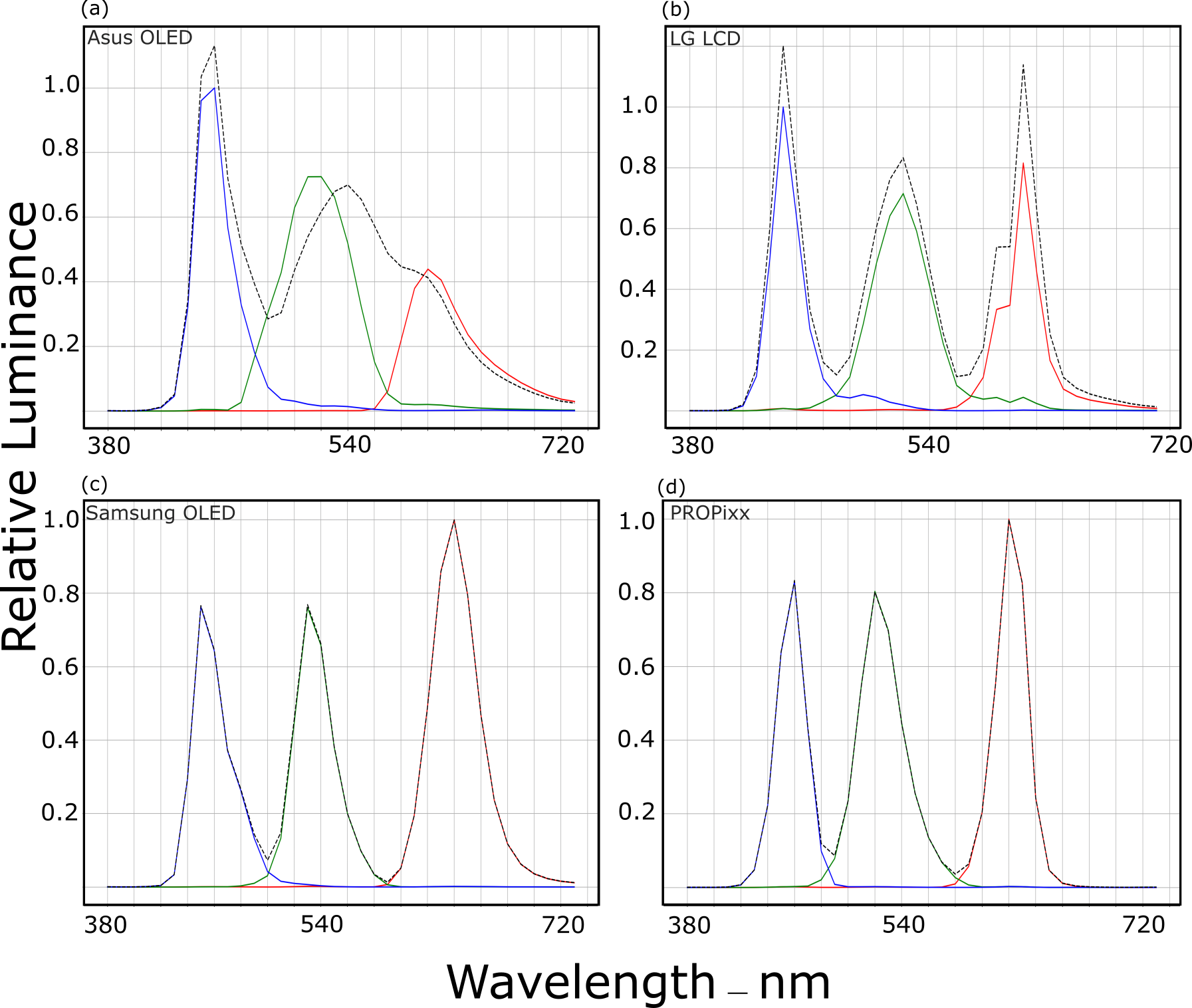}
    \caption{Spectral power distribution (SPD) curves of the 4 tested displays. The colors of the curves correspond to the measured channels RGB accordingly. In addition to the black curve representing the white color SPD. a) Asus OLED, b) LG LCD, c) Samsung OLED TV, and d) PROPixx. This arrangement of sub-figures is followed throughout this paper in this order. The vertical tick spacing on the x-axis corresponds to a step of 20 nm.}
    \label{fig:display_SPDs}
\end{figure}

\subsection{Spectral Power Distribution - SPD}

In Fig.\ref{fig:display_SPDs}, the four display's SPDs are plotted side-by-side for each of their primaries (R, G, and B) channels plus a dashed curve representing the white SPD. Narrower curves and less overlap are favorable as that indicates more reproducible color variations. Also if a system has a perfect additive property one should notice that by looking at how the white SPD behaves in an RGB 3-primary system. That is, the white is a result of the activation of all used primaries - talking strictly about 3-primary systems- then the white spectral curve would overlap perfectly with the separate RGB channels' curves, which is the case for both the Samsung OLED TV and the PROPixx projector, Fig.\ref{fig:display_SPDs}(c) and (d) respectively. For LG LCD the white SPD seems to have a perfect overlap over its primaries however a little bit amplified in magnitude. Whereas for Asus OLED, it is clear that the white SPD does not look like its the exact summation of its RGB primaries, Fig.\ref{fig:display_SPDs}(a), keeping in mind that this Asus OLED model uses a 4-primary system RGBW. One of the purposes of having a white subpixel in such systems is to increase the display's brightness range. In addition to that the additional white subpixel plays a role in reducing power consumption (i.e. it can cut power consumption by half compared to RGB-subpixel systems).
\begin{figure}[!h]
    \centering
    \includegraphics[width=0.8\textwidth]{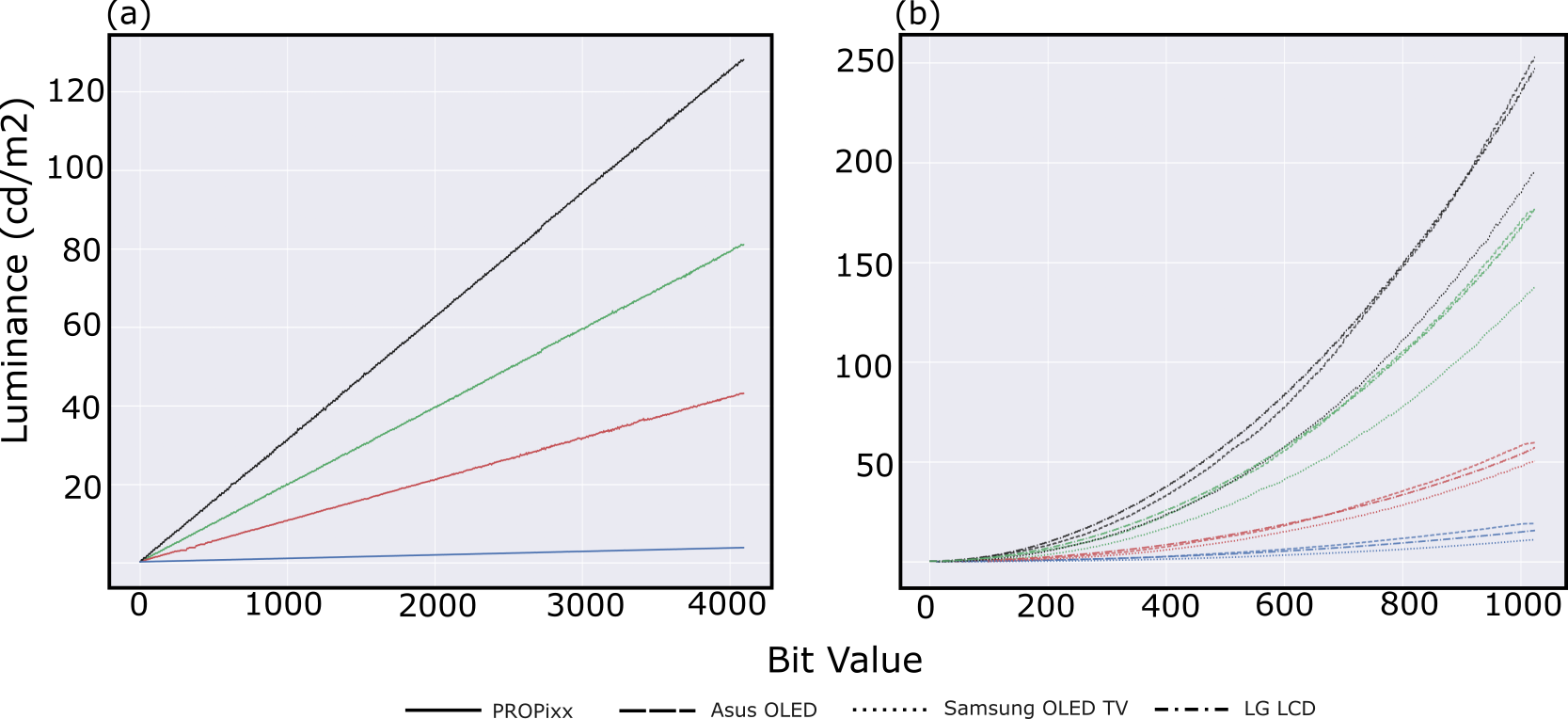}
    \caption{Luminance ramps of RGB channels plus grayscale of the 4 display systems, a) PROPixx, b) contains Asus OLED (dashed), LG LCD (dash-dot), and Samsung OLED (dotted). The curves represent the measured channels RGB with their corresponding colors, Grayscale is in black. The 3 display systems (Asus OLED, Samsung OLED TV, LG LCD) have a similar behavior and do not show any major difference in this test. The PROPixx projector is perfectly linear as expected. Pay attention to the x-axis that is not unified.}
    \label{fig:RGBGray_ramps_all}
\end{figure}
Also, it enlarges the color gamut and hence represents more pure colors \citep{OLED_White.ch8}. It is worth noting that \citet{Chestermann2016}, in their work, found heuristically that driving the pixels to their white point with maximum pixel value triplet (255, 255, 255 in an 8-bit system) activates not only the white subpixel but also the red and the blue subpixels as well. However, they found experimentally a triplet of (252, 255, 215) that would activate only the white subpixel.
For why the Samsung OLED TV behaves differently from Asus OLED, we know that Asus OLED monitor uses the underlying LG LW270AHQ-ERG2 panel which is RGBW. In contrast, Samsung OLED TV uses Quantum Dot (QD-OLED) which mainly relies on a blue back-light with quantum dots layer and red and green filters to output RGB colors. The quantum-dots technology contributes to enhancing the display's primaries and hence having narrow-band spectral curves for the primaries, hence a larger color gamut as one can observe by comparing the curves of Samsung vs. Asus OLED systems, Fig.\ref{fig:display_SPDs}(c) vs. (a) respectively.

\subsection{Luminance Response}

\begin{figure}[!h]
    \centering
    \includegraphics[width=0.8\textwidth]{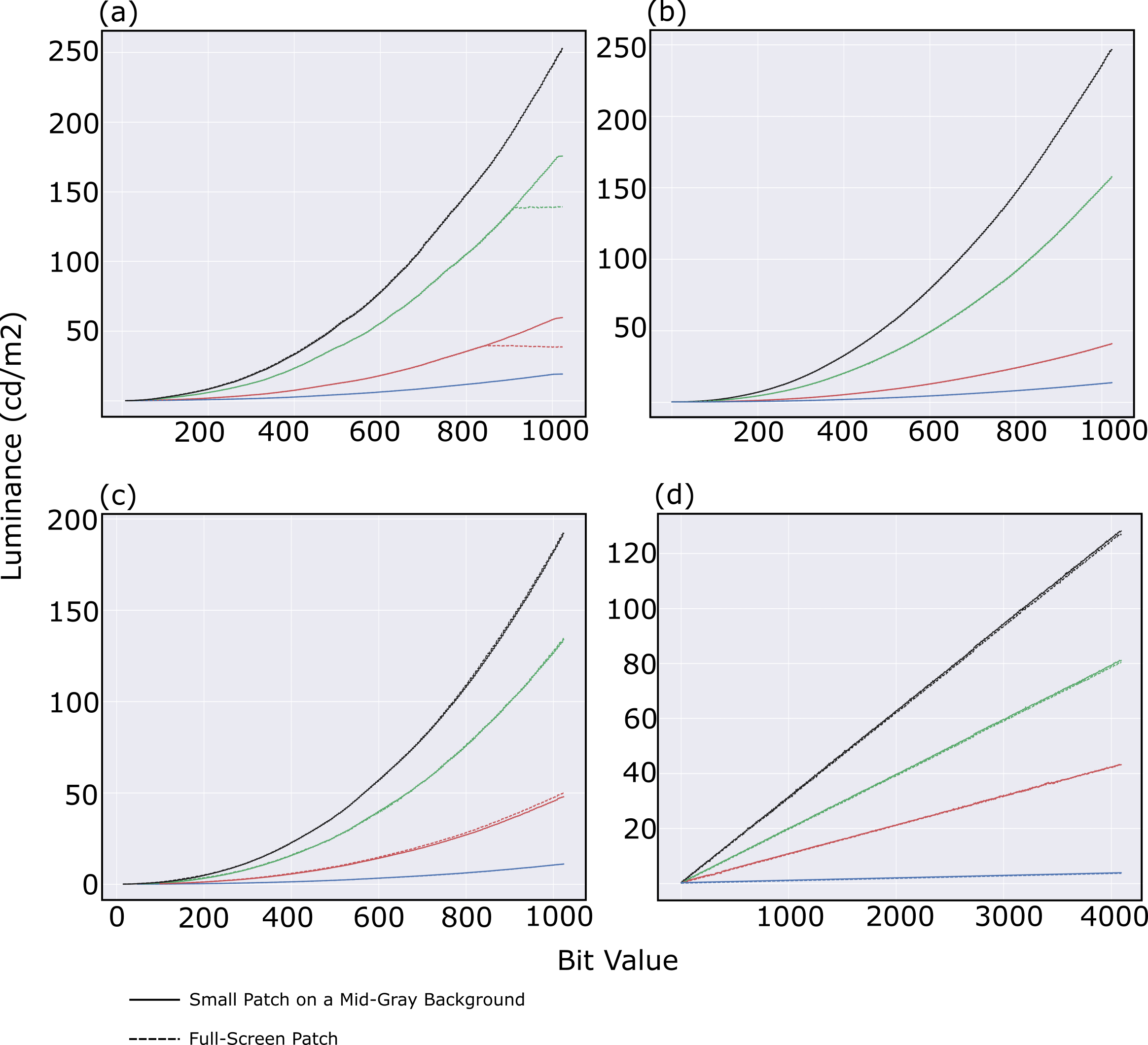}
    \caption{Luminance ramps of RGB channels plus grayscale of the 4 display systems, a) Asus OLED, b) LG LCD, c) Samsung OLED TV, and d) PROPixx. The luminance curves show 2 different tests, solid lines represent small color patch ($512\times512$) on a full-screen mid-gray background versus dashed-lines represent full-screen color patch behaviors. Note the unequal x- and y-axes between the panels.}
    \label{fig:fullscreen_vs_small_patch_RGBGray_ramps}
\end{figure}

\begin{figure}[!h]
    \centering
    \includegraphics[width=0.75\textwidth]{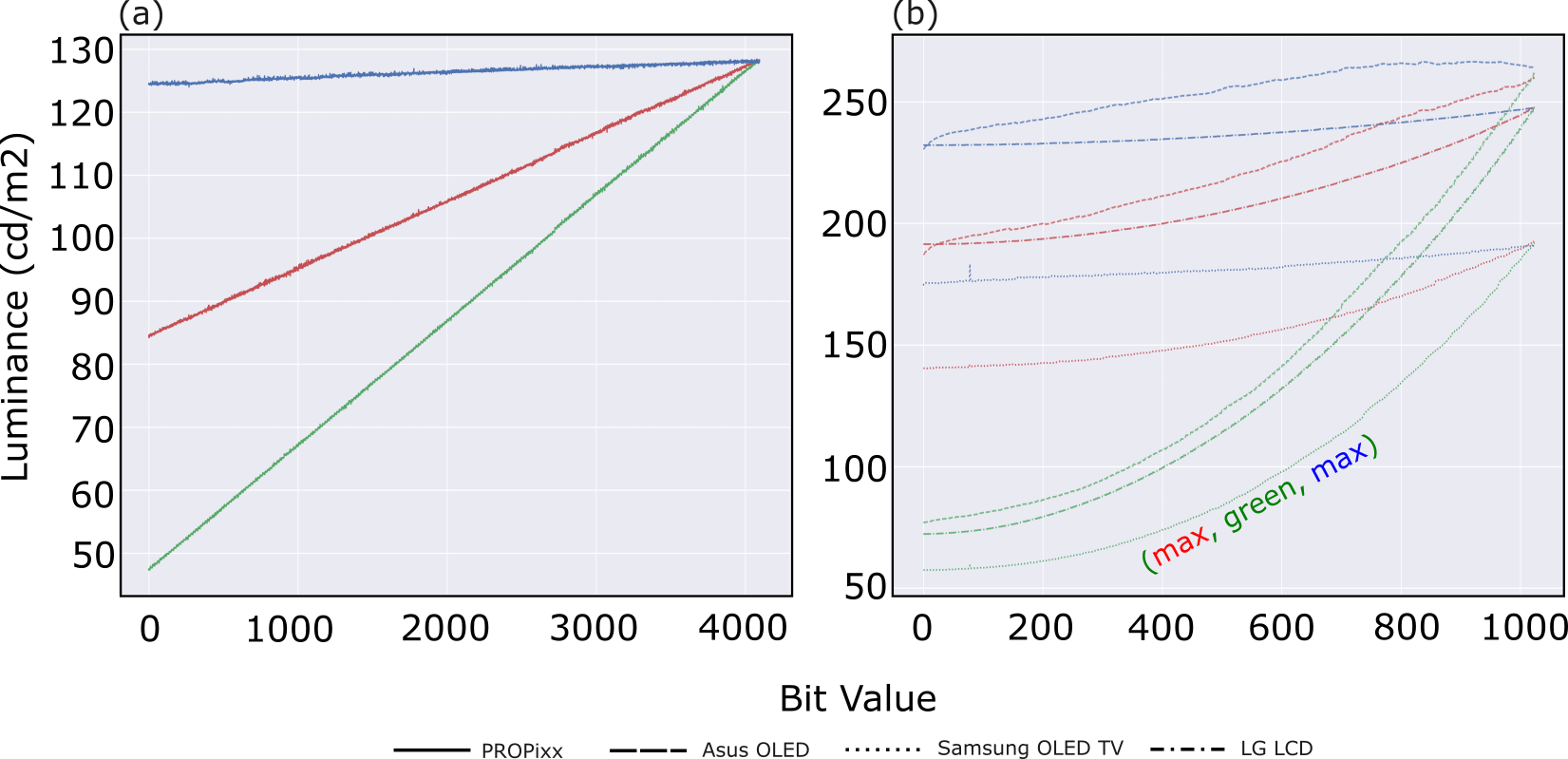}
    \caption{Luminance ramps of RGB channels, when 2 channels are saturated and one is ramping up, for ASUS OLED monitor (dashed), SAMSUNG OLED TV (dotted), and LG LCD monitor (dash-dot). The colors of the curves match the channel that is varying in value.}
    \label{fig:ASUS_LG_SAM_RGB_OCS}
\end{figure}

Fig.\ref{fig:RGBGray_ramps_all} shows the luminance ramps of the RGB primaries and grayscale. The measurements of Asus OLED are plotted in dashed lines, the LG LCD's are in dash-dotted lines and the Samsung OLED TV's are in dotted lines --Fig.\ref{fig:RGBGray_ramps_all}(b), while the measurements for PROPixx are plotted separately on the left sub-figure Fig.\ref{fig:RGBGray_ramps_all}(a). The colors of the curves match the measurements of the corresponding RGB channels and the black curves for the grayscale measurements. The general behavior of all 3 displays (Asus OLED, Samsung OLED TV, and LG LCD) seems to be very similar to one another. PROPixx is calibrated to behave in a linear fashion by default as can be seen from Fig.\ref{fig:RGBGray_ramps_all}(a). Changing the background color from mid-gray to black did not show any sign of difference that was worth the attention on the tested displays. It is worth mentioning here that in \citet{Ashraf24} discussing a different OLED model, namely LG 27" UltraGear 27GR95QE which is also an RGBW system, their measurements of a small patch on a black background showed a sign of saturation for all channels at a certain bit value, nearly $ >700$, in the upper half of the 10-bit range.

We repeated the test however this time with the color patch filling the display, full-screen color, activating all the display's pixels to act in synchrony. Fig.\ref{fig:fullscreen_vs_small_patch_RGBGray_ramps} shows the results in 4 subfigures one for each display system, a) Asus OLED, b) LG LCD, c) Samsung OLED TV, and d) PROPixx projector. The curves in the figure take on the same color as the measured channel, in addition to black for the grayscale. The solid lines show the same test result as in Fig.\ref{fig:RGBGray_ramps_all} (small color patch on a mid-gray full-screen) while the dashed lines are for the full-screen color patch measurements. Asus OLED monitor stands out among all 4 tested displays for its behavior in full-screen mode -- In Fig.\ref{fig:fullscreen_vs_small_patch_RGBGray_ramps}(a) the green channel starts to saturate at a luminance level near $138\; cd/m^2$ and pixel value of $916$ in a 10-bit range in comparison to the maximum green channel luminance for a small patch at $175.5 \; cd/m^2$ , and the red channel starts to saturate around luminance level of $\approx 39\; cd/m^2$ around pixel value of $849$ in comparison to maximum luminance level of $59.6$ for a small color patch. The blue channel's and the grayscale's behaviors stay unaffected\footnote{We noticed that if we adjust some of the 6-axis saturation settings, the overall display saturation goes down and the effect of ceiling gets resolved, however on the expense of other metrics like the additivity breaking down}. The saturation behavior seems to agree with the findings of \citet{Chestermann2016} for the same type of OLED (RGBW) that states that when all the pixels are active then luminance saturation of RGBW OLED is inevitable at higher bit value -- also known as digital driving level (DDL). Both the LG LCD and the
Samsung OLED TV use backlight but completely different technologies for producing RGB subpixels, and neither of them shows any sign of limitation similar to Asus OLED for reaching the maximum luminance while driving all the display's pixels at once -- full-screen mode. PROPixx projector, as well, does not suffer any limitation in this regard.

While measuring OCS, we noticed that for ASUS OLED the maximum luminance exceeds the maximum luminance level when measured at the end of the grayscale ramp. The blue channel is very noticeably overreaching the calibrated maximum luminance, which is $250\; cd/m^2$ as can be seen from the results in dashed lines for the blue channel in Fig.\ref{fig:ASUS_LG_SAM_RGB_OCS}(b).
This should be understood in the light of \citet{Chestermann2016} study, for in RGBW OLED systems the maximum white is usually not a result of just mixing the common 3 primaries RGB, nor just activating the white subpixel by itself. Rather, it is a combination of white and other subpixels in action so that the display outputs more luminance. On the other hand, for Samsung OLED TV (dotted curves), the OCS behavior is more steady except for one abrupt spike that occurs around a bit value of $77$ during the measurements, we could not explain why this happened even in repeated measurements. The spike happens across all channels and is the strongest in the blue channel which amounts to a change in the luminance level from $\approx 176$ to  $183\; cd/m^2$ a change of nearly $4\%$. Note also the maximum luminance level for Samsung OLED TV of $\approx 192\;cd/m^2$ in comparison to LG LCD and Asus OLED that is $\approx 250\; cd/m^2$. PROPixx, Fig.\ref{fig:ASUS_LG_SAM_RGB_OCS}(a) behaves very linearly as expected. LG LCD and PROPixx did not show any behavior that was worth the attention.

\subsection{Channel Additivity}

\begin{figure}[!h]
    \centering
    \includegraphics{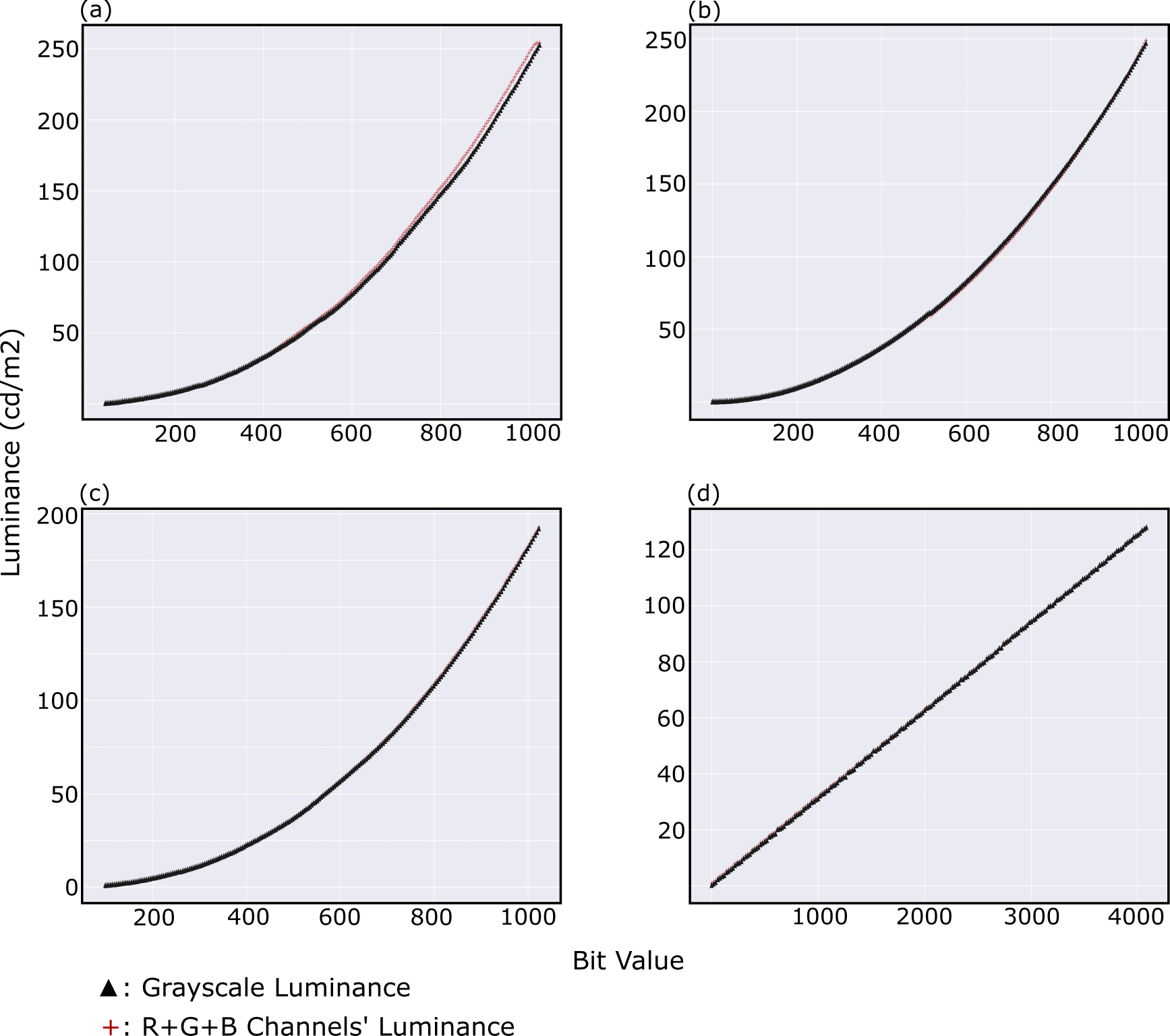}
    \caption[width=0.6\textwidth]{Additivity test showing the luminance levels of the summed RGB primaries at a certain bit value ('+' sign) and the corresponding luminance level for the same bit value along the grayscale ('$\blacktriangle$' sign).}
    \label{fig:Additivity_summation}
\end{figure}

Table \ref{tab:additivity_RGB_White} shows the exact values for the white luminance versus the summed maximum luminance of the individual RGB channels showing that a linear behavior governs all four display systems under the test. 
Fig.\ref{fig:Additivity_summation} illustrates the additive behavior all along a grayscale (black to white) in which the $\blacktriangle$ sign represents the luminance level of a grayscale value (i.e. subpixels acting together), while the '+' sign represents the summation of the luminance levels at that value when the RGB channels were measured individually. All four display systems show a congruent behavior in which the RGB channels act individually in the same way as when they act collectively to produce a certain luminance level under the chosen settings. One notices in Fig.\ref{fig:Additivity_summation}(a) for Asus OLED, 10-bit, that in the upper half, the summed values do not align perfectly with the grayscale luminance values as is the case for the rest. However, after looking into the statistics of this deviation, it showed that the maximum this deviation reaches is $\leq 2\%$ -- which should not raise any major concerns.

\begin{table}[!h]
    \centering
    \caption{Additivity test checks whether the maximum luminance of  $R_{max}+G_{max}+B_{max} \stackrel{?}{=} White$.}
    \label{tab:additivity_RGB_White}
\begin{tabular}{|l|l|l|}
\hline
\thead{Display / Luminance (cd/m2)} & \thead{R+G+B} & \thead{Max White} \\ \hline
Asus OLED & 254.3 & 252.7   \\  \hline
Samsung OLED TV & 192.7 & 192.1    \\  \hline
LG UltraGear & 249.1 & 247.1    \\  \hline
PROPixx & 128.1 & 128.1    \\  \hline

\end{tabular}
\end{table}

\subsection{Linearity and Gamma function ($\gamma$)}
After setting everything to user-mode/manual, adjusting to the aforementioned settings, and running color calibration, we found the following gamma values and summarized them in Table \ref{tab:displays_gamma} for the different displays. All displays were calibrated for $\gamma = 2.2$, however upon fitting a power function into their various output luminance curves, a deviation in the encoded gamma values has come to light and was more noticeable in certain displays than in others. For instance, Samsung OLED TV shows the least deviation of its gamma values across its RGB channels and the grayscale ($\leq 0.01$). On the other hand, Asus OLED display shows a small yet noticeable difference between its Red-channel gamma value and the rest of the channels which amounts to a maximum difference of $\approx 0.1$. The LG LCD display gamma values across its channels show a deviation of no more than $0.09$ at most between the green channel and the blue channel. PROPixx, as it was mentioned before, comes with perfect linear behavior that governs all its channels combined or individually.

\begin{table}[!h]
    \centering
    \caption{Gamma curve fitting of RGB channels and grayscale of each tested display after color calibration.}
    \label{tab:displays_gamma}
\begin{tabular}{|l|l|l|l|l|}
\hline
\thead{Display / Gamma ( $\gamma$ )} & \thead{R} & \thead{G} & \thead{B}& \thead{Grayscale} \\ \hline
Asus OLED & 2.24 & 2.17 & 2.14 & 2.18   \\  \hline
Samsung OLED TV & 2.30 & 2.29 & 2.29 & 2.29    \\  \hline
LG UltraGear & 2.04 & 2.07 & 1.98 & 2.01    \\  \hline
PROPixx & 1.0 & 1.0 & 1.0 & 1.0    \\  \hline

\end{tabular}
\end{table}

\subsection{Color Gamut}
\begin{figure}
    \centering
    \includegraphics[width=0.45\textwidth]{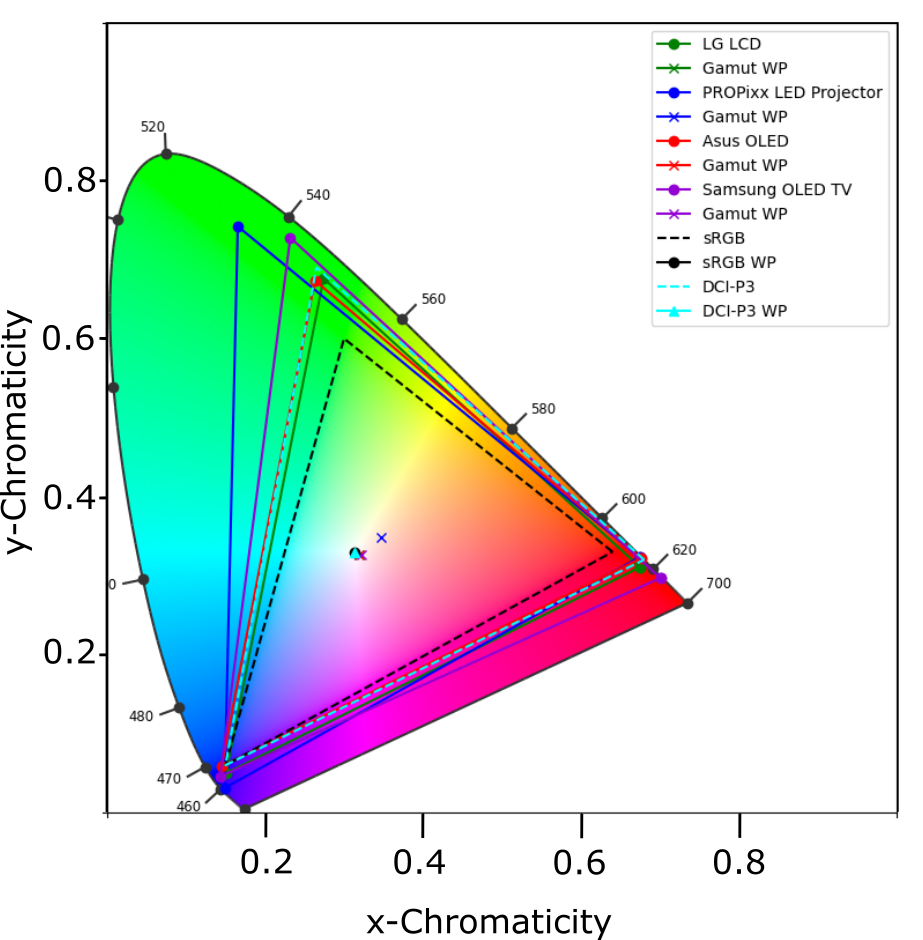}
    \caption{Color gamut of the 4 tested display systems represented on CIE 1931 xy-Chromaticity diagram with their corresponding white-points (WP). sRGB and DCI-P3 color gamut are plotted as references. Note that the PROPixx in High Bit-Depth mode has not been calibrated to D65, unlike the other monitors.}
    \label{fig:ASUS_LG_SAM_PROPixx_color_gamut}
\end{figure}

A representation of each color gamut can be seen in Fig.\ref{fig:ASUS_LG_SAM_PROPixx_color_gamut} along their white-points (WP); precise chromaticities are provided in Table \ref{tab:diplays_xy_Chromaticities}. 
In the Fig. \ref{fig:ASUS_LG_SAM_PROPixx_color_gamut} the common small \textit{sRGB} and the wider \textit{DCI-P3} color spaces are represented in dashed lines as references. Both Asus OLED and LG LCD approximate very closely the DCI-P3 color gamut. Whereas, Samsung OLED shows a larger color gamut thanks to its quantum-dots technology and the narrow (pure) primaries it can achieve. PROPixx also shows yet a larger color gamut, hence a richer color experience, thanks to the optimized selection of its RGB LED primaries. 
All white-points seem to align with the D65 WP, except for PROPixx which seems to be a little bit warmer and lies around D50.
This behaviour likely has two sources: first, we were not able to calibrate the color profile of the PROPixx as for other monitors using our calibration software.
Second, our measurements are made in the PROPixx's "High Bit-Depth" mode, which results in slightly cooler LEDs than the default RGB 120~Hz mode and therefore a shifted white-point (Peter April, personal communication).
One would be able to calibrate a D65 white-point for the PROPixx in High Bit-Depth mode using a custom LUT if desired.

\begin{table}[h!]
    \centering
    \caption{xy-Chromaticity coordinates of displays' primaries RGB and white-point. Note that the PROPixx in High Bit-Depth mode has not been calibrated to D65, unlike the other monitors.}
    \label{tab:diplays_xy_Chromaticities}
\begin{tabular}{|l|l|l|l|l|l|l|l|l|}
\hline
\thead{Display / xy-Chromaticity } & \multicolumn{2}{c|}{\thead{Red}} & \multicolumn{2}{c|}{\thead{Green}} & \multicolumn{2}{c|}{\thead{Blue}} & \multicolumn{2}{c|}{\thead{White-Point}} \\ \hline
\thead{ -- } & \thead{x} & \thead{y} & \thead{x} & \thead{y} & \thead{x} & \thead{y} & \thead{x} & \thead{y} \\ \hline
Asus OLED & 0.676 & 0.322 & 0.263 & 0.673 & 0.145 & 0.0582 & 0.319 & 0.327  \\  \hline
LG UltraGear & 0.675 & 0.311 & 0.274 & 0.676 & 0.151 & 0.050 & 0.319 & 0.327\\  \hline
PROPixx & 0.674 & 0.323 & 0.166 & 0.742 & 0.149 & 0.031 & 0.346 & 0.348   \\  \hline
Samsung OLED TV & 0.700 & 0.297 & 0.232 & 0.727 & 0.143 & 0.045 & 0.319 & 0.323  \\  \hline
\end{tabular}
\end{table}

\subsection{Luminance Uniformity}

\begin{figure}[!h]
    \centering
    \includegraphics[width=0.7\textwidth]{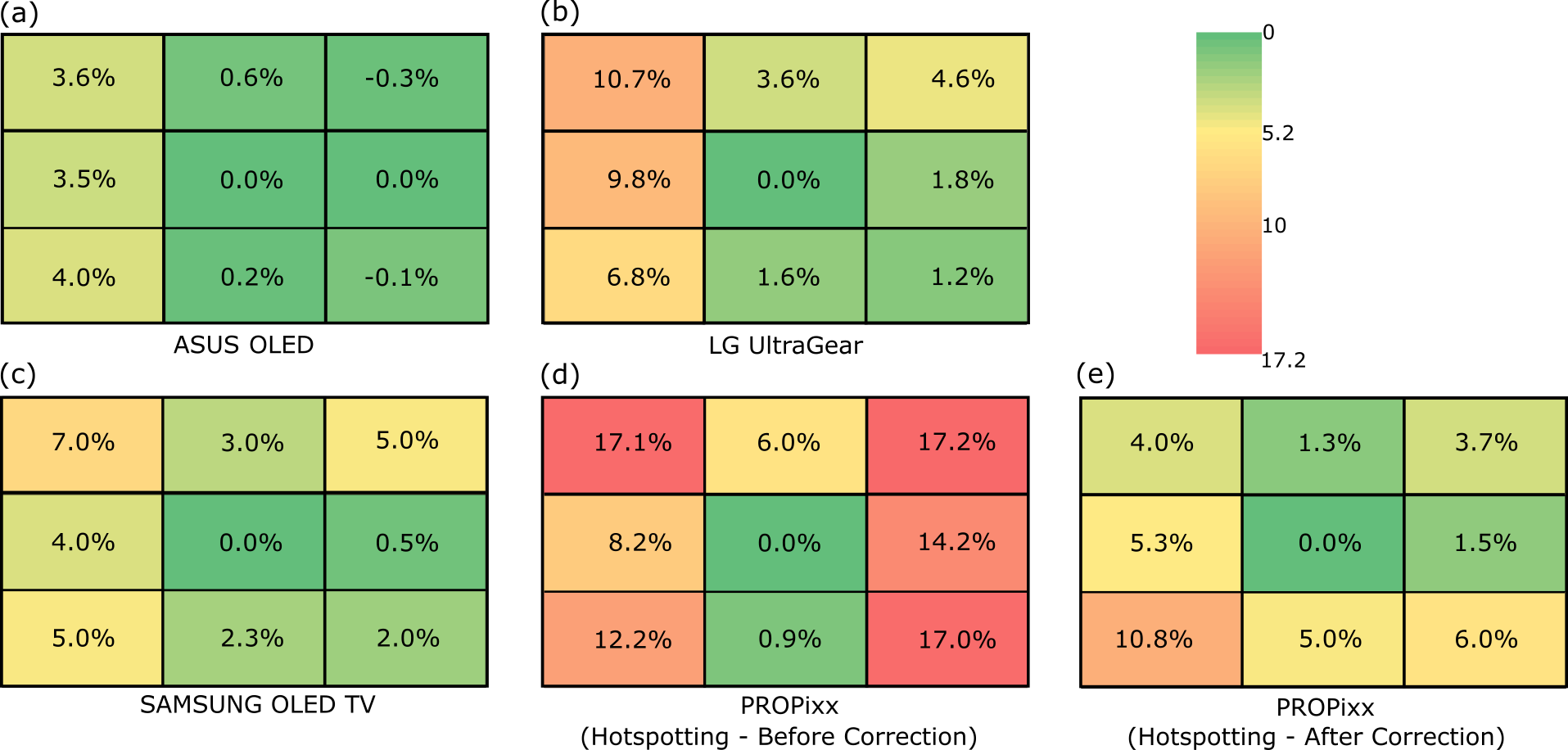}
    \caption{Luminance uniformity across each of the tested displays. The display is divided into 3X3 grid and luminance was measured at the center of each. The difference is reported as an average percentage relative to the central part of each display. Negative values indicate that the luminance level at that part of the display is higher than the central part. The heatmap is universal across all the displays and in absolute value.}
    \label{fig:ASUS_LG_SAM_PROPixx_uniformity}
\end{figure}

Luminance uniformity across a display ensures that presenting a stimulus with a fixed luminance level at different parts of the display does not result in a dimmer or brighter representation depending on its location. We divided the display into a $3\times3$ grid (i.e. 9 sections) and measured the luminance ramp of a grayscale in each section individually. Fig. \ref{fig:ASUS_LG_SAM_PROPixx_uniformity} shows the average difference between the 8 sections of each display relative to its central part -- which is usually considered to be the most stable part of any display system. The negative numbers indicate that higher luminance levels of that part of the display than the central part were registered.

The LG LCD display shows a higher average difference than the other two displays Asus and Samsung OLED across its surface with a difference as high as $10.7\%$ in the worst-case scenario -- It seems that the left half of the LG LCD display performs far worse than the right half. Both OLEDs have maximum differences that amount to $\approx 4\%$ and $\approx 7\%$ for  Asus and Samsung respectively. The PROPixx projector suffers from what is known as \textit{hotspotting}, which is a well-known problem in projectors and causes strong variations in luminance across the projection area, in our case up to $17.1\%$ difference. VPixx Technologies offers a calibration procedure in which the effect of hot-spotting can be mitigated. Fig.\ref{fig:ASUS_LG_SAM_PROPixx_uniformity}(e) shows the result of luminance difference after running hotspotting correction in which the luminance uniformity across the display has improved. However, the left-bottom corner still shows a maximum luminance difference as big as $10.8\%$. It is worth mentioning that after the correction, the maximum luminance level measured at the center dropped from $\approx 128$ to $97\; cd/m^2$. Out of the 4 tested displays, Asus OLED seems to show the most luminance uniformity across its surface.

\subsection{Filling Factor}
We measured the luminance output of a rectangular stimulus presented in the middle of the display starting at dimensions of $10\%$ of the display resolution and going up to full-screen $100\%$ with an increment of $10\%$. The background color was set to one of three options: black, white, or mid-gray. The stimulus color was either one of the primaries (R, G, or B) or grayscale (black to white), and depending on the available bit depth ranging from either $[0 - 1023]$ for $10$-bit displays with a step of $50$ or $[0 - 4095]$ for the $12$-bit PROPixx with a step of $100$.

\begin{figure}[!h]
    \centering
    \includegraphics[width=0.7\textwidth]{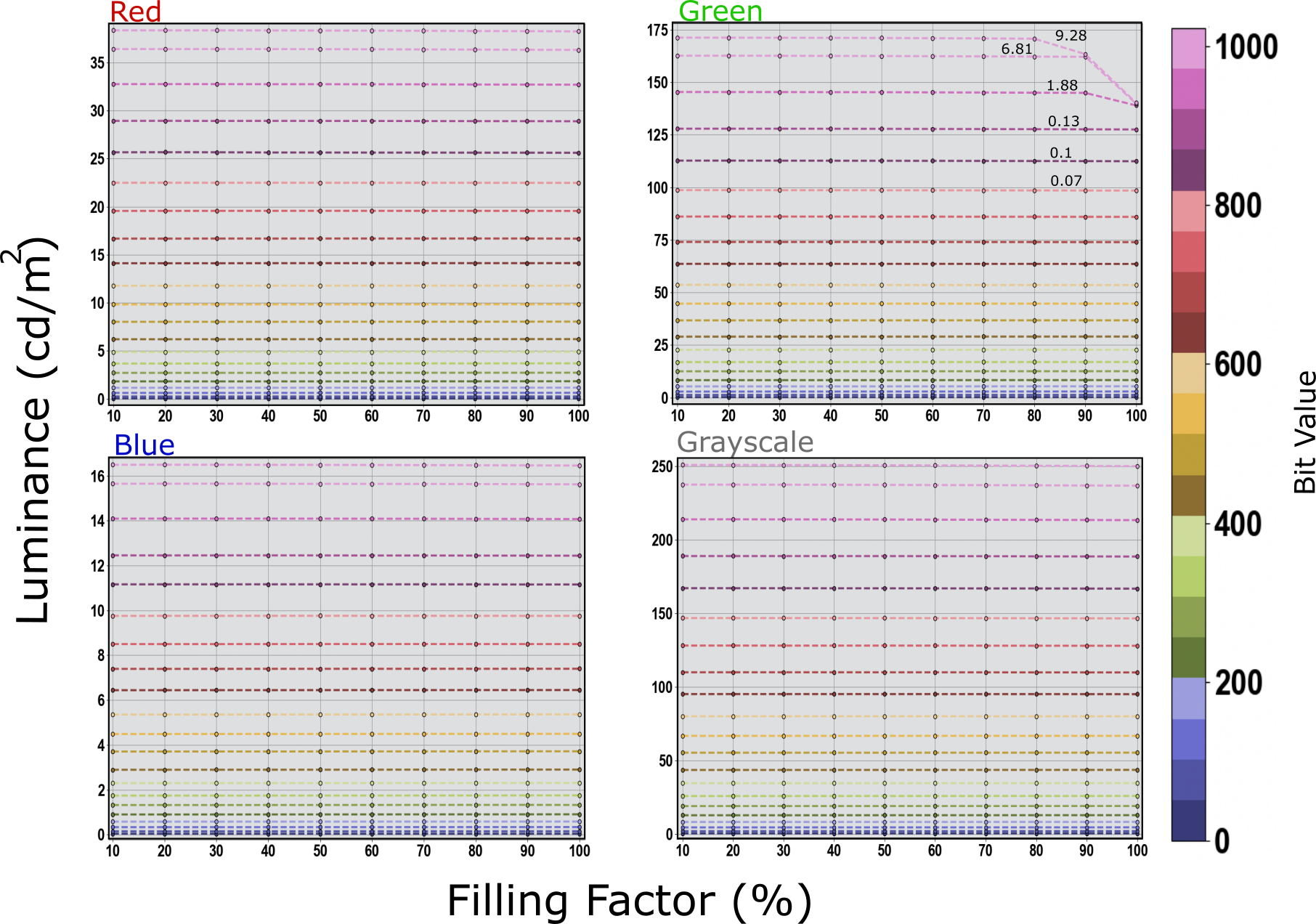}
    \caption{Luminance output as a function of filling factor (percentage of the stimulus area) for different bit values $[0 - 1023]$ with a step of $50$. The filling factor influence (aka. Average Pixel Level) is tested on Red, Green, Blue channels and the grayscale (black to white). The background color is set to mid-gray. All the channels show steady luminance output independent of the filling factor except for the upper range of the bit values, $>900$, of the green channel where it shows a standard deviation up to $9.28$ and a drop in luminance from $\approx 171.3$ to $\approx 140.3\; cd/m^2$. A few standard deviation values are written on top of the last four tested bit values.}
    \label{fig:ASUS_FF}
\end{figure}

\begin{figure}[!h]
    \centering
    \includegraphics[width=0.7\textwidth]{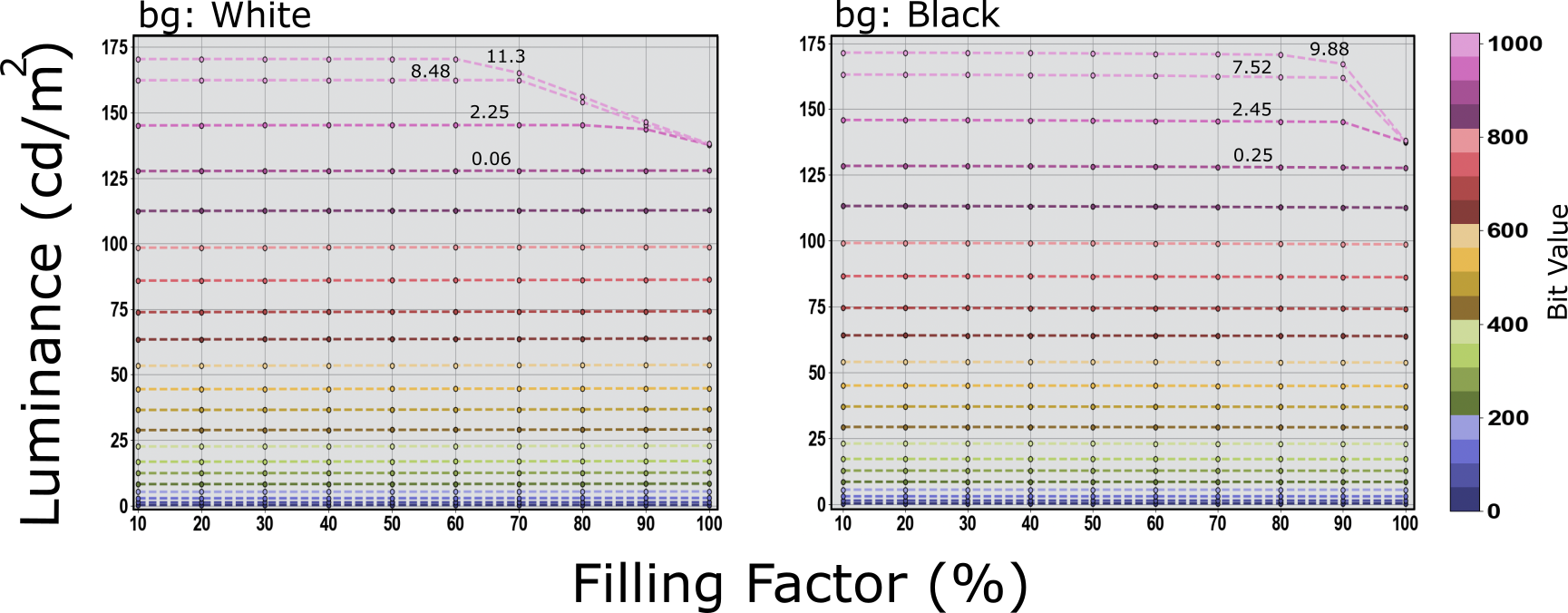}
    \caption{The effect of the background color on the luminance output of the Asus OLED Green channel. A white background (left) affects the luminance output at higher bit values causing a standard deviation across the various filling factors of $11.3$ and luminance change from $\approx 170$ to $\approx 138\; cd/m^2$ starting from a filling factor $60\%$ up to $100\%$. The black background (right) shows a maximum standard deviation of $9.88$ and luminance change from $\approx 171.5$ to $\approx 137.9\;cd/m^2$.}
    \label{fig:ASUS_FF_bg}
\end{figure}

The measurements for the three display systems, Samsung OLED TV, LG LCD UltraGear, and PROPixx projector, did not show any luminance dependency on the filling factor or the background color. The maximum reported standard deviations across all the measurements of R, G, B, and grayscale and different filling factors were found to be $0.97, 0.12, 0.94$ for the Samsung, LG, and PROPixx respectively when the background is set to mid-gray. Changing the background color did not show any sign of larger deviation or abnormal luminance fluctuation worth the attention. For the Asus OLED, though, a deviation in the luminance behavior was observed only for the green channel and only at a very high bit value $>900$ in $10$-bit depth. A drop in luminance starts to kick in after a filling factor $>90\%$ and after $80\%$ for a bit value $>950$ and $>1000$ respectively when the background is either black or mid-gray, check Fig.\ref{fig:ASUS_FF}(Green) and Fig.\ref{fig:ASUS_FF_bg}(bg:Black). When the background is rather white (all the background pixels are at their maximum) a drop in luminance starts to be noticeable at a bit value of $>=950$ and a filling factor $>70\%$, and as early as a filling factor of $60\%$ for the maximum bit value $1023$, check Fig.\ref{fig:ASUS_FF_bg}(bg:White). The standard deviations across different filling factors for the green channels with mid-gray background, specifically for the last six tested values $[800, 850, 900, 950, 1000, 1023]$ are $[0.07, 0.1, 0.13, 1.88, 6.81, 9.28]$ respectively as can be read on the green channel plot Fig.\ref{fig:ASUS_FF}(Green). Only a few standard deviation values are written on top of the last four lines that correspond to the tested values for the green channel with different background as they are the most affected values out of all the other measurements. The maximum standard deviation of $9.28$ is the result of a maximum drop in luminance under the testing settings from $\approx 171.3$ to $\approx 140.3\; cd/m^2$. When changing the background color, only the green channel kept showing a drop in luminance and only for bit values $>900$. For the white background (all the background pixels at their maximum) the standard deviation reaches $11.3$ for the maximum bit value $1023$ and shows a drop in luminance from $\approx 170$ to $\approx 138\; cd/m^2$ across different filling factors. Surprisingly, when the background is set to black (all background pixels are off) the maximum standard deviation is still comparable to when the background is set to mid-gray and amounts to $9.88$ and the change in luminance across the various filling factors goes from $\approx 171.5$ to $\approx 137.9\;cd/m^2$.
We refrain from plotting the rest of the measurements for the other display systems because of the little deviations they show as they seem to keep their luminance output behavior steadier and constant regardless of the filling factor and the background color.

\subsection{Pixel Response Time and Waveform}
\begin{figure}[!h]
    \centering
    \includegraphics[width=0.8\textwidth]{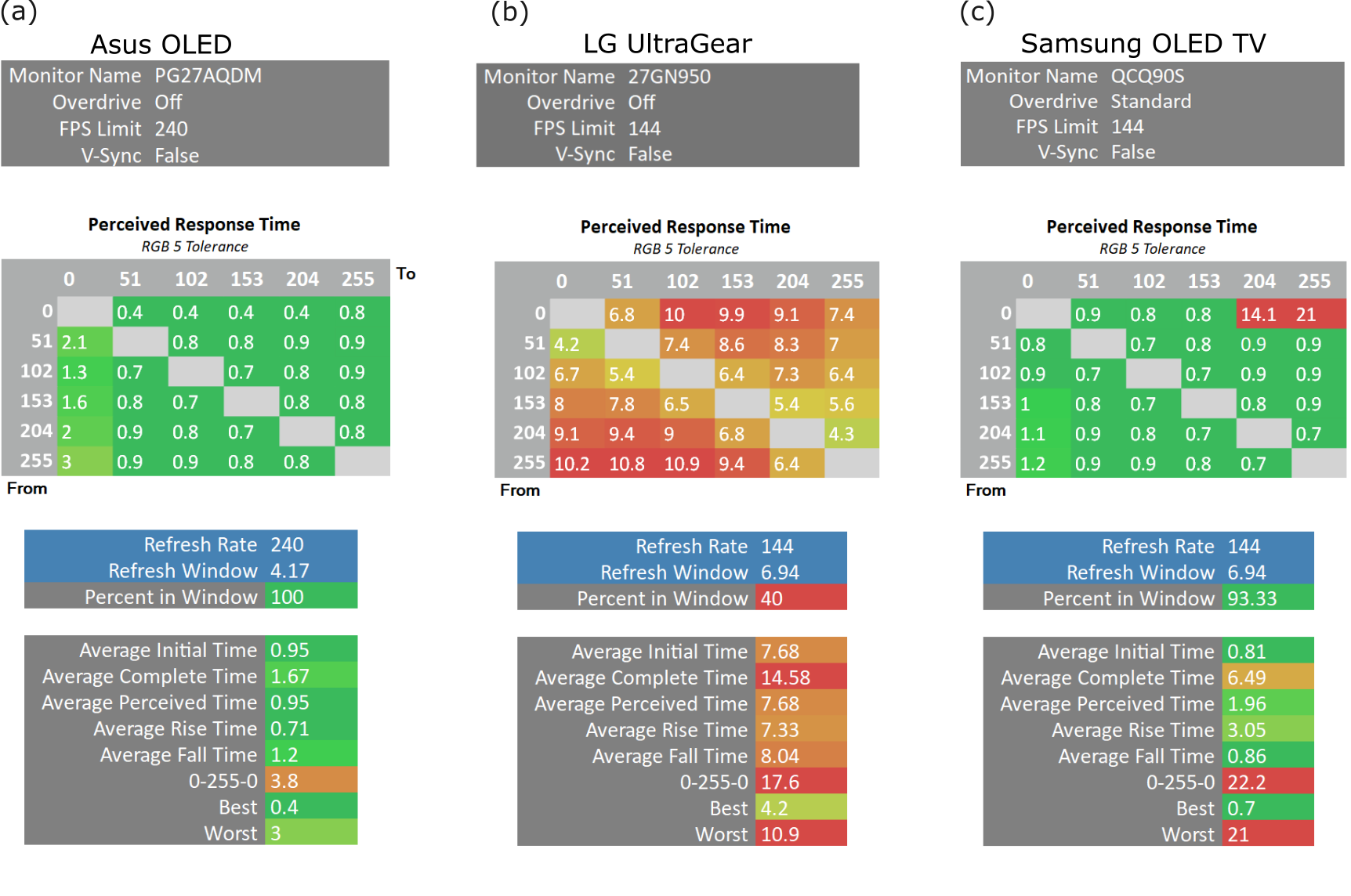}
    \caption{The transition time of pixel responses (Gray-to-Gray; G2G) between 6 uniformally-spaced gray levels reported in milliseconds with a tolerance level of 5 RGB values for the 3 display systems Asus and Samsung OLED, and the LG UltraGear LCD using OSRTT PRO CS.}
    \label{fig:response_box}
\end{figure}

Pixel response is how long it takes a pixel or a group of pixels to make a transition from one state to another to output different colors or light levels. The most extreme transition would happen when the two-pixel values are very far apart i.e. black to white or vice versa. OSRTT response box (RB) allows us to measure how long it takes on average for a full-screen color to make a transition in what is known as Gray-to-Gray transition i.e. All pixel transitions happen in the grayscale range [0 - 255] after scaling the actual bit depth of the selected display down from e.g. 10-bit to 8-bit. The step between every two consecutive gray colors is uniform and calculated by dividing the whole range into 6 parts i.e. a step is $255/5 = 51$ plus the black ($0$). 
Fig.\ref{fig:response_box} shows the response time in milliseconds for the 3 display systems, a) Asus OLED, b) LG LCD, and c) Samsung OLED TV, each with their corresponding native refresh rate as indicated under \textit{"Refresh Rate"}. Each set of measurements contains 4 sub-tables, top to bottom they are:
1) The table on the top contains information about the display model under the test, whether overdrive is activated, the maximum tested FPS (Frame Per Second) limit, and whether V-Sync is activated if available. 
2) The 2nd table contains the transition time in milliseconds between every pair of gray colors (G2G) expressed as \textit{"Perceived Response Time"}. 
3) The 3rd table contains information about the detected settings of the display such as the refresh rate, the actual time of the refresh window in milliseconds, and the test window size (100 for full-screen). 
Finally, 4) the table at the bottom contains summary information about the measurements regarding the "Average Initial Time" which is the metric recommended for measuring the response time -- that is between $10\%$ and $90\%$ of the signal ignoring any over- or undershoot if any occurred \citep{ICDM.p164}. The "Average Complete Time", that is until the signal reaches effectively the target value and stabilizes. The "Average Perceived Time", that is measuring the signal until it is within the pre-defined tolerance including any over- or undershoot if it happens. Then the "Average Rise Time" and "Average Fall Time" express the rising and the falling time of the signal from one state to another in milliseconds. "0-255-0" is the total transition from black to white and back to black. "Best" and "Worst" are self-explanatory for the best and the worst transition time in this set of measurements.

\textit{"Perceived Response Time"} differs from the "Initial Response Time" measurements for it keeps measuring the signal up until after any overshoot or undershoot if either was part of the signal. In other words, until it settles close to the end target state within a pre-defined tolerance -- here the tolerance is set to 5 RGB bit values. The smaller the transition time the more control the display has over its pixel states, hence offering more readiness for an accurate change on-demand with little to no delay. We will focus on "Perceived Response Time" in this report for it is more realistic and relatable to the human experience in vision science experiments.

Looking at the measurement of the LG LCD in Fig.\ref{fig:response_box}(b) shows the well-known issue of LCD IPS that requires on average $7.68\;ms$ to make a change to the liquid crystal from one state to another, in line with other findings \citep[e.g., ][]{Elze2012TemporalPO}. 
Whereas Asus OLED shows a very fast response time, Fig.\ref{fig:response_box}(a), with an average of $0.95\;ms$ and worst is $3\;ms$ which is still below the time required for displaying 1 frame at its maximum refresh rate $\frac{1000\;ms}{240\;Hz}=4.1\bar6$. On the other hand, Samsung OLED TV shows an average transition time of $1.96\;ms$ and the worst is $21\;ms$ for the transitions toward white $255$. While its signal fall-time looks very reasonable for an OLED display, which is $0.86\;ms$, its rise-time on the other hand looks a bit sluggish with $3.05\;ms$ on average when compared to Asus with $1.2 \; ms$ and $0.71\; ms$ respectively. At the time of writing the manuscript, we could not justify why only the transition to the gray level $204, 255$ in Samsung OLED TV accounts for too much delay in the signal in comparison to other gray levels and to the Asus OLED display in general.

For PROPixx this test was not possible to run as the OSRTT software could not communicate with the projector setup. Hence, the corresponding data could not be measured.

\subsubsection{Waveform}
We recorded measurements of a single frame during a transition from black to white to black (0-255-0) in three ways. First, with a static flashing stimulus (i.e. a $512\times512$ white box in the middle of the display in full-screen mode with a mid-gray background) for exactly one frame every second for at least 10 seconds. Second, the same flashing stimulus but this time displaying it for a period of 11 frames every second. Finally, the same stimulus translated across the display from left to right at a speed of 50 pixels per frame.
Doing so allows us to measure closely the time it takes to process one frame, to see if that is influenced by a longer presentation of a stimulus or a moving stimulus, and to measure the signal's waveform more closely. This part of the experiment was coded in \textit{Psychopy}, with careful attention paid to minimizing and removing dropped frames. 
We used OSRTT RB hardware and software to record the signal using its "live mode".

\begin{figure}
    \centering
    \includegraphics{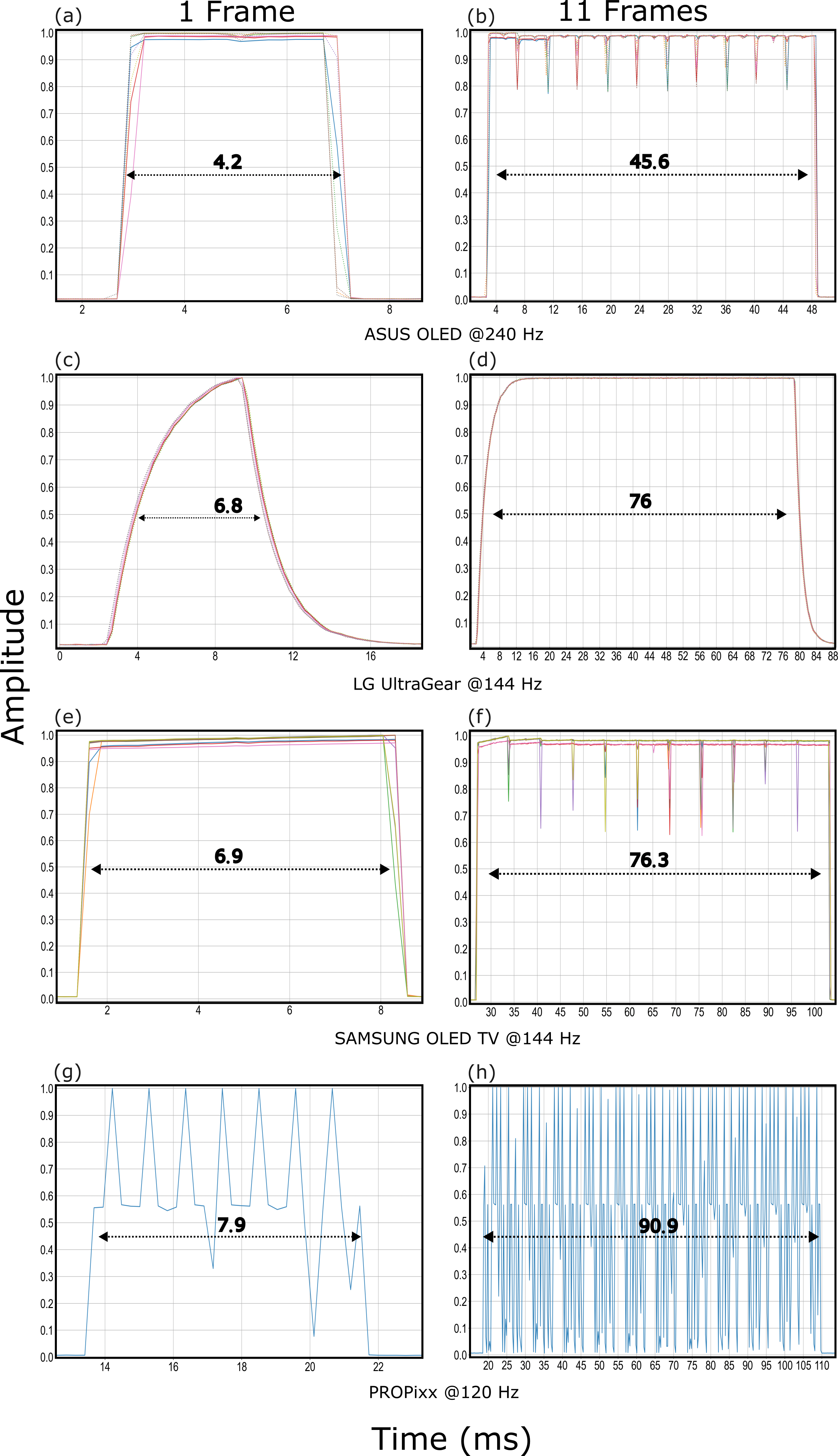}
    \caption{Waveform measurements for 1 frame and 11 frames a second at the native refresh rate of each corresponding display system. 9 measurements in total superimposed on top of one another for each except for PROPixx measurements where this was avoided due to the jittery waveform that would make the figure very cluttered otherwise. The unit of the measurements is milliseconds, an average approximation of the 9 measurements is written on each sub-figure. Note that the x-axis scale differs across subfigures.}
    \label{fig:waveform}
\end{figure}

Fig.\ref{fig:waveform} shows the waveforms of pixel responses for 1 and 11 frames for the 4 display systems and the average duration it takes for this transition to happen averaged across 9 measurements in total for each. The durations are calculated at the signal's full-width-half-maximum (FWHM) i.e. at $50\%$. As expected, the plots show a sharp rise and fall of the signal for the OLED and PROPixx DLP projector with a very accurate duration when presenting the specified stimulus. E.g. Samsung OLED TV can be driven at a maximum refresh rate of $144\; Hz$ a second, hence 1 frame should last $\frac{1000 \; ms}{144 \; Hz}= 6.9\bar4\;ms \rightarrow$ 11 frames would last $\approx76.3\bar8 \; ms$. Even though the LG LCD shows an accurate duration of presenting the stimulus for a certain number of frames at its FWHM, it is clear from its waveform that the presented stimulus reaches its desired luminance level (1.0) only for a brief period for the liquid crystal requires a considerable amount of time to change its state which is evident in the signal rising and falling shapes. In contrast, the OLED and DLP technologies show instantaneous rise and fall signal response, hence the stimulus is presented for exactly all the duration of a complete frame at its desirable luminance level. The DLP PROPixx projector waveform looks very spiky compared to all other systems due to how the micromirrors in DLP systems operate\footnote{A curious reader may consult \citep{TexasInstr.Ch1} for how digital micromirror devices (DMD) work}.

For a moving stimulus, the response box (photodiodes) was placed horizontally while the stimulus moved left to right several times. Fig.\ref{fig:MovingStim} shows the response signal as the stimulus moves across the sensors, hence the progressive growing and diminishing of the amplitude. The average signal duration is given on each subplot accordingly given the refresh rate the display was operating at. For both OLED displays, there is no observable change in the signal response behavior, while for LG LCD the response varies while the stimulus moves across the sensors ranging from $6.6$ to $7.5 \; ms$ at $144 \; Hz$. PROPixx projector display shows a little deviation from its theoretical value of $\frac{1000\;ms}{120\; Hz}= 8.\bar3\;ms$ that amounts to a maximum of $0.5 \; ms$.

\section{Discussion}
The metrics we used in this report to assess the quality and suitability of different displaying systems show that it is hard for a consumer-level display system to meet all the stringent requirements a vision scientist may demand for various experiments. Even within the same model, variations and inconsistencies are unsurprising, since the primary focus of consumer-level displays is not precise and accurate physical display but rather for daily activities such as movie watching, gaming, and similar tasks. Consumer-level display technology, nonetheless, has come a long way since CRT displays became the vision science gold standard. Here, we have tested four of the recent and most advanced displaying systems to evaluate their suitability for vision science. Three systems are consumer-level and relatively budget-friendly, while the other system is considered a professional scientific device, namely the PROPixx projector.

\begin{figure}[!h]
    \centering
    \includegraphics[width=0.7\textwidth]{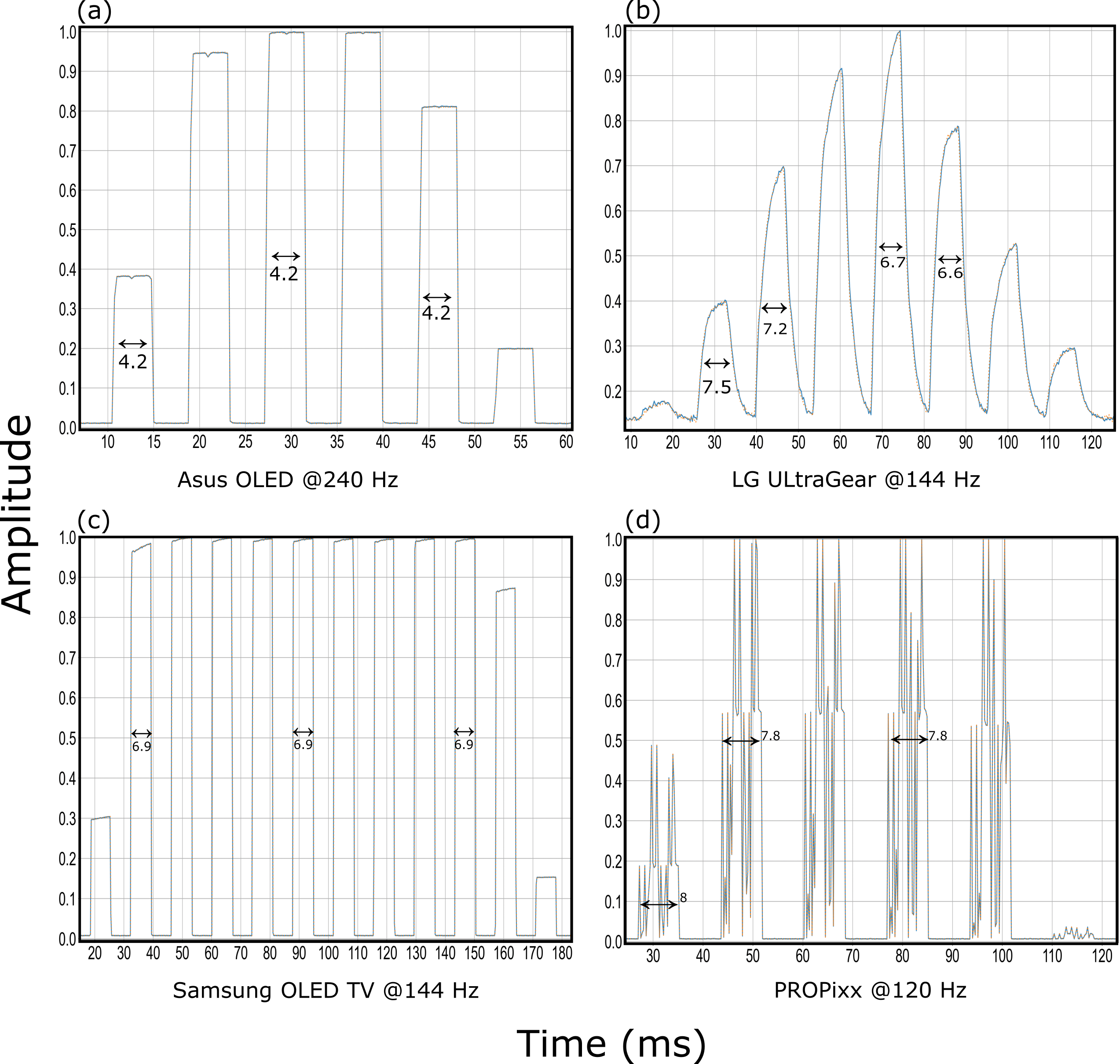}
    \caption{Measuring a moving stimulus (rectangle) horizontally from left to right across each display. The photodiode sensors are also placed horizontally, hence the waveform response's progressive growing/diminishing amplitude. Note that the x-axis differs between sub-figures.}
    \label{fig:MovingStim}
\end{figure}

Testing the behaviors of the luminance curves and their encoded gammas, after careful selection of settings for each display and after color calibration, showed that all three consumer-level displays (Asus OLED, Samsung OLED TV, and LG LCD) follow a predictable gamma power function that enables an experimenter to decode the gamma and linearize their output if desired. The gamma values of each of the RGB channels and the grayscale luminance curves are presented in Table.\ref{tab:displays_gamma} and can be seen visually in Fig.\ref{fig:RGBGray_ramps_all}.
As \citet{Cooper2013OLED} concluded in their paper, it is possible to have luminance output following a power law (gamma function) upon careful settings selection so that the monitor output is linearizable. Our color calibration process did not affect the luminance monotonic ramps, at least not for the small-patch mode, rendering the whole bit depth to be useable unlike the findings in \citet{Ashraf24} paper where after performing color calibration ceiling/saturation starts to kick in around a bit value of $\approx 700$ in a 10-bit depth in the small-patch condition. 

The measured area for each display is a $512 \times 512$ in the center while the background is set to mid-gray unless otherwise is mentioned. During our tests, we did not notice any difference in the luminance behavior if the background changed from mid-gray to black. However, changing the color patch size from $512 \times 512$ to full-screen patch size resulted in a change in the luminance behavior only for the Asus OLED display system and only for the red and green channels, where these channels started to saturated at bit values of $\approx 849$ and $916$ respectively in a 10-bit depth range. Whereas for all the other displaying systems, no change in their luminance behavior has been recorded, showing an improvement in hardware behavior compared to the tested hardware in \citet{Elze_OLED_2013}'s study, in which saturation in full-screen mode was an inevitable characteristic for their tested hardware.

In another test, we checked if maxing up two of the three RGB primaries would influence the behavior of the third primary by saturating two channels and ramping up the remaining one. The results are plotted in Fig.\ref{fig:ASUS_LG_SAM_RGB_OCS}. In the case of Asus OLED and because it relies on 4-subpixel primaries, RGBW, one notices that the blue channels, when the red and the green are saturated, start to exceed the maximum recorded luminance level that of the calibration, and exceeds the maximum measured luminance of its grayscale ramp. We understand this behavior in the light of \citet{Chestermann2016}'s study that pointed out that in RGBW systems usually representing white color activates not only the white subpixel nor only the combined maximum of the three RGB primaries but rather a combination between the white subpixels and other subpixels as well. Given that, we suspect that while fixing the red and the green channels to the maximum and ramping up the blue channel to reach the white color, along the bit range the white subpixel started to be active and act together with the changing subpixel, hence the overreaching in the luminance level. Samsung OLED, on the other hand, did not show a similar behavior due to its use of a different technology (Quantum-Dot).

In the additivity test, all four display systems showed a very nice behavior in which their individual RGB channels summed up to the corresponding luminance level when they act combined to reproduce grayscales. The results were summarized in Table \ref{tab:additivity_RGB_White} and visualized in Fig.\ref{fig:Additivity_summation}. These findings are consistent with the measurements of \citet{Cooper2013OLED} and \citet{Ito2013OLED}, in which their measured OLEDs showed a nice additivity behavior with a maximum difference in luminance levels between ($R_{max}+G_{max}+B_{max}$) and $white$ to be $<1.0\%$.

Measurement of the color gamut across the four display systems revealed that the PROPixx, followed by the Samsung OLED TV, exhibited the widest color gamut (Fig.\ref{fig:display_SPDs}). In comparison, the Asus OLED and LG LCD systems covered up to $\approx 94\%$ of the DCI-P3 color space. A larger color gamut allows a richer color experience, more saturated colors, and more color variations. The color gamut is a result of the selection of the underlying primaries, the narrower their spectral curves the larger the color gamut they define.
All the displays' color gamut are larger than the standard \textit{sRGB} and some even larger than \textit{Adobe RGB 1998} making them very suitable options for experiments requiring a wide range of colors. 

The luminance uniformity test showed that Asus OLED has the least luminance variation across its display surface  (Fig.\ref{fig:ASUS_LG_SAM_PROPixx_uniformity}(a)), when dividing the display into a $3\times3$ grid, measuring grayscales and comparing them against the central part of the display/grid. 
All three consumer-level display systems show worse average differences on the left half of the display, Fig.\ref{fig:ASUS_LG_SAM_PROPixx_uniformity}(a,b and c), which we assume is due to the electronics concentration on that side. 
The PROPixx projector suffers from a phenomenon called \textit{hotspotting} in which it exhibits a highly non-uniform luminance behavior across its display surface. After calibrating and correcting for hotspotting, the variation in luminance improved from a maximum average difference of $17.1\%$ (before) $\rightarrow 10.8\%$ (after). The left-bottom corner of the projector display did not show much improvement before and after the correction in comparison to the other parts of the grid (Fig.\ref{fig:ASUS_LG_SAM_PROPixx_uniformity}(d) vs. (e)). It is important to remember that the properties of PROPixx are a function of the projected image size and distance. i.e. the hotspotting effect, among other characteristics, would differ depending on these parameters.
To the best of our knowledge, there are only a few luminance uniformity measurements for OLED displays in the literature to this date. \citet{Ito2013OLED}, for example, has measured the uniformity of Sony PVM-2541 OLED, the same model used in \citet{Cooper2013OLED} study, and found little to no difference in luminance output across the display surface when measured from a distance and along the surface normal, the maximum reported difference is $< 3\%$. Our measurements show a difference of maximum $\approx 7\%$ for the Samsung OLED TV and maximum $\approx 4\%$ for the Asus OLED gaming monitor given that the  displays are color-calibrated.

The measurements of the average pixel level (APL) or as mentioned here the filling factor show no sign of abnormal or sudden change in luminance output that could depend on the stimulus size or the background color for the three displaying systems (Samsung OLED TV, LG LCD UltraGear, and the PROPixx projector). All these three systems were invariant to the filling factor (stimulus size) and the background color (black, white, or mid-gray) with a maximum standard deviation of $0.97, 0.12, 0.94$ for the Samsung, LG, and PROPixx respectively when the background is set to mid-gray. However, only the green channel of Asus OLED gaming display shows a sign of unsteady luminance output, under the selected test conditions, and only for bit values $>900$ in a $10$-bit range and a filling factor $>60\%$. The Asus green channel suffers the most when the background is set to white (all background pixels at their maximum) causing a maximum standard deviation of $11.3$ caused by the change in luminance across the various filling factors, starting from filling factor $>60$, from $\approx 170$ to $\approx 138 \;cd/m^2$, check Fig.\ref{fig:ASUS_FF_bg}(bg:White). The black and the mid-gray background both show standard deviation values of $9.88$ and $9.28$ respectively. Surprisingly, even when all the background pixels are off (black), this does not seem to help regulate the luminance output of the green channel better. Notably, the common way of testing APL is by targeting the grayscale range (black to white) regardless of the behavior of the individual RGB channels. Our APL grayscale measurements show very robust behavior in luminance output regardless of the background color (black, white, or mid-gray) with a maximum standard deviation of $0.27$. The unexpected luminance change happening in the green channel should not be a practical limitation but more of a technical issue under certain conditions. As long as the green channel is not being used under extreme conditions, e.g. filling factor $>60\%$ and bit value $>900$, this should not impose any practical limitation on the display performance. These findings show an immense improvement from the reported results in previous reports such in \cite{Ito2013OLED, Tian_OLED_HDR, Yang_OLED_vs_LED}.

Regarding the pixel response time, consistent with earlier reports \citep{Ghodrati2015TheO, Cooper2013OLED, Elze2012TemporalPO}, we find that the tested LCD display exhibits sluggish behaviour (Fig.\ref{fig:response_box}(b)). On the other hand, both OLED devices show a very quick response, as little as $0.3, 0.4\;ms$ (Fig.\ref{fig:response_box}(a)\&(c)). 
\citet{Ito2013OLED}'s analysis suggests a similar or slightly slower transitional duration of $\approx 2\; ms$ to reach a target luminance level. 
However, the Samsung OLED TV performs poorly for the transition between gray values of $204, 255$, with sluggish response that goes up to $21\;ms$ at $144\;Hz$, which is far beyond the theoretical duration of a single frame $\frac{1000\;ms}{144\;Hz}=6.9\bar{4}$. 
During the time of writing the manuscript, it remained unclear to us why this behavior keeps repeating only for a transition towards gray values of $204, 255$. 
With this exception, based on our measurements it seems that OLED response time is still fast enough and accurate for, virtually, any perception experiment.

The Asus OLED proved to be very accurate and precise in its response time behavior, for its slowest response would take only $3\;ms$, which is still under its theoretical frame duration $\frac{1000\;ms}{240\;Hz}=4.1\bar{6}$ (Fig.\ref{fig:response_box}(a)). 
Analyzing closely the waveform of the pixel response (Fig.\ref{fig:waveform}) shows the instantaneous and sharp rise and fall of OLED signal for Asus and Samsung OLED, which enables presenting the desired stimulus at its target luminance level for virtually the complete period of one or more frames as requested. 
On the other hand, by observing the waveform of LG LCD pixel response, one sees how presenting a stimulus for exactly 1 frame would not faithfully present it at its desired luminance level for the whole frame duration but rather only for a brief period, being affected by the gradual rise and fall of the signal as a result of turning the liquid crystals that eat up a considerable amount of time of the total frame duration \citep[Fig.\ref{fig:waveform}(c); see also ][]{Elze2012TemporalPO}. 
The PROPixx waveform is also sharp and square, however the spikes one can observe in the recorded data is due to the photodiode interacting with the DLP flipping micro-mirror inside the projector. 
These spikes should not have any effect on the target luminance level. While capturing a moving stimulus that moves from left to right across the display, the Asus OLED, Samsung OLED TV and PROPixx projector, all have sharp, clear, and instantaneous signal (Fig.\ref{fig:MovingStim}). In contrast, the LG LCD shows a variation in the presented signals of the moving stimulus for it varies between $6.6 \rightarrow 7.5 \;ms$ for a stimulus that is supposed to have a duration of only $\frac{1000\;ms}{144\;Hz}=6.9\bar{4}\;ms$ (Fig.\ref{fig:MovingStim}(b)).

Contemporaneous results to ours have recently been published as a preprint by \citet{dimigenHighspeedOLEDMonitor2024}. 
Specifically, they evaluated the same ASUS ROG Swift OLED PG27AQDM that we do in this paper, and compared this to two LCD monitors (an ASUS and an Iiyama) and an Iiyama CRT monitor.
Compared to our measurements, their paper focused more on the temporal properties of monochromatic stimuli.
They included a number of measurements that we did not make, including dependence on viewing angle and warmup time (temperature).
In addition, they included a test of high-frequency flicker stimulation and an intra-saccadic stimulation experiment.
In contrast, our paper more thoroughly characterises the long-presentation luminance and color properties of this monitor.

Similar to our measurements, they find that the ASUS OLED spatial luminance uniformity does not exceed $5.5\%$ in any two locations they measured across the display surface (grid of $3\times5$). 
Even though they followed a different technique to ours, the spatial uniformity deviation is very close to one another ($max\; 5.5 \;vs.\; max\;4.0\; \%$). 
The transition times they recorded for the signal rise and fall are $\approx 0.3\;ms$. 
Our evaluation of the signal rise and fall time were slightly different with $\approx 0.7$ and $1.2\; ms$ respectively. 
Nonetheless, the full width at half maximum measurements of the signal duration of one frame align very well in both papers with one another conforming with the expected nominal value at $240\; Hz$. 
\citet{dimigenHighspeedOLEDMonitor2024} show how activating "Uniform Brightness" helps in having more controlled Auto-Brightness Limiting (ABL) behavior and limiting the occurrence of dimming depending on the display content. 
However, they find that this option is only of use under Windows OS, while testing it on Linux OS showed some hardware intervention changing the luminance output when display brightness is set to $100\%$. 
They advise, hence, when using Linux OS to limit brightness to only $40\%$ ($\approx 140\;cd/m^2$). 
Our results instead suggest that under Linux OS it is still possible to make use of the "Uniform Brightness" option. 
We did not see this drastic drop in luminance in relation to the filling factor (Average Picture Level) when the display brightness is still $80\%$ (calibrated for $\approx 250\;cd/m^2$) when testing the whole range of a grayscale (black to white). 
Nonetheless, our APL measurements for the grayscale still align with \citet{dimigenHighspeedOLEDMonitor2024} by showing a robust luminance output independent of the APL percentage or the background color (black, white, or mid-gray).
In general, the similarity of these results from two different laboratories using different control hardware and measurement equipment should instill confidence in the reader that the overall good performance of the ASUS OLED we report here is unlikely to be accidental.

\subsection{Caveats and limitations}

It is important to note that the measurements and findings we present are specific to the tested display units under the selected settings and calibration methods. 
As a result, similar performance across identical models cannot be assumed or guaranteed. 
Informally, we observe that the behavior of the displays is significantly influenced by the chosen settings. 
While we took significant effort to select settings that produced the best possible performance for each monitor over a variety of stimulus aspects, we cannot guarantee that better (or worse) performance could not be found, nor can we guarantee that different exemplars of the same display devices will perform similarly.

All our luminance measurements were taken after running color calibration on the corresponding systems, except for PROPixx, which would affect the luminance output and the behavior of the RGB primaries. 
To our knowledge, all related literature relies rather on the default display profile while only tweaking the preset options for color spaces and gamma by dialing it directly on the display hardware. 
In our experience, the hardware preset and the default display profile would not reflect faithfully and accurately the selected values and can suffer from inaccuracies. 
We advise doing color calibration and profiling so that the display profile is manually and accurately tuned to the desired peak luminance, and the behavior of the display primaries is appropriately corrected for the desired gamma and luminance output, besides having accurate color reproduction if one desires. 
One has, as well, the option to calibrate only for luminance output, i.e. grayscale, and omitting color accuracy if color is not of any importance to the target experiment.

\section{Conclusion}

We find that at least two consumer-level OLED display devices may be suitable for vision science experiments (holding important caveats for each). 
The Asus OLED ROG Swift PG27AQDM performed very well across our test battery, with the exception that its red and green channels saturate in full-screen mode measurements, and the green channel showed changes in luminance output as a function of stimulus size at high intensities. 
The Samsung OLED TV also performed very well, except for the pixel response time test, in which it performed very poorly for only certain G2G transitions. 
The professional PROPixx projector system performed very well in all tests, except for spatial uniformity due to hotspotting, but remains the best over all monitors we tested (albeit with a price tag some 20 times higher than the Asus).
In summary, consumer-level display systems seem to have improved since previous reports and offer opportunities for vision scientists to use high-performance systems for relatively little cost compared to professional displays.

\section{Acknowledgements}
Co-funded by the European Union (ERC, SEGMENT, 101086774, to TW). Views and opinions expressed are however those of the author(s) only and do not necessarily reflect those of the European Union or the European Research Council. Neither the European Union nor the granting authority can be held responsible for them. 
We thank Weiliang Wang for assistance with Samsung OLED data collection, and Julian Klabes for providing a conversion script to verify the luminance measurements with the Konica Minolta CS2000. 
We also thank Uli Wannek, Peter April, Lindsey Fraser and the VPIXX scientific staff for their feedback on the manuscript.

\section{Contributions}
\textbf{TAH:} Conceptualization, coding, data measurements, and analysis, preparing, writing, and editing the manuscript. \textbf{KK:} Data collection for Samsung OLED TV and reviewing the manuscript, \textbf{TQK:} Reviewing the manuscript. \textbf{TW:} Conceptualization, project administration, editing and reviewing the manuscript.

The measurements are publicly available at https://doi.org/10.5281/zenodo.13970348

\newpage

\bibliography{jovtemplate_latex}
\bibliographystyle{jovcite}

\newpage

\section{Appendix: Monitor settings}\label{ap:display_settings}

This appendix describes the settings used for each of the consumer monitors.

For the \textbf{ASUS OLED} display we have adjusted the display settings to the following: 
Brightness: 96, 
Contrast: 80, 
Uniform Brightness: On, 
    Color Space: DCI-P3,
    Color Temperature [R, G, B]: [100, 97, 82], 
    Gamma: 2.2, 
    Game Visual: User Mode,
    Screen Saver: Off,
    Power Setting: Standard Mode,
    Lighting Effect: Off,
    Variable Refresh Rate: Off,
    Shadow Boost: Off.

For the \textbf{LG UltraGear LCD} we adjusted the display settings to the following:
    Game Mode: Gamer1, Response Time: Off, Adaptive Sync: Off,
    Brightness: 42, Contrast: 80, Sharpness: 80, Color Temperature [R, G, B]: [43, 40, 39],
    Gamma: Mode 1, Energy Saving: off, refresh rate: 144 Hz, Dimming: Off 

For the \textbf{Samsung OLED TV} the settings were set to the following:
    brightness: 50 (max), Contrast: 50 (max), Sharpness: 10, Color: 25, Color Tone (G/R): 0, Color tone: Standard, Shadow-Tracing: 0, Color Space Settings: Normal, Gamma: 2.2, White-Balance $\rightarrow$ 2 points $\rightarrow$ RGB Gains: [37, 18, -31], RGB Offsets: [0, 0, 0].

\end{document}